\documentstyle{ar209}

\renewcommand{\evensidemargin}{0pc}
\renewcommand{\oddsidemargin}{0pc}
\def\lsim{\mathrel{\rlap{\lower4pt\hbox{\hskip1pt$\sim$}}
    \raise1pt\hbox{$<$}}}         
\def\gsim{\mathrel{\rlap{\lower4pt\hbox{\hskip1pt$\sim$}}
    \raise1pt\hbox{$>$}}}         
\begin{document}

\input epsf.tex    

\input psfig.sty

\jname{Annu. Rev. Nuclear and Particle Science}
\jyear{2001}
\jvol{1}
\ARinfo{1056-8700/97/0610-00}

\title{Atomic Parity Nonconservation and Nuclear Anapole Moments}

\markboth{Haxton \& Wieman}{Anapole Moments}

\author{W. C. Haxton
\affiliation{Institute for Nuclear Theory, Box 351550, and Department
of Physics, \\
University of Washington, Seattle, WA 98195, and \\
Department of Physics, University of California, Berkeley, CA 94720}
C. E. Wieman
\affiliation{JILA and \\
Department of Physics, University of Colorado, Boulder, CO 80309-0440}}
\begin{keywords}
anapole moment, atomic parity nonconservation, radiative corrections, weak interactions
\end{keywords}

\begin{abstract}
Anapole moments are parity-odd, time-reversal-even moments of
the E1 projection of the electromagnetic current.
Although it was recognized, soon after the discovery of parity
violation in the weak interaction, that elementary particles
and composite systems like nuclei must have anapole moments,
it proved difficult to isolate this weak radiative correction.
The first successful measurement, an extraction of the nuclear
anapole moment of $^{133}$Cs from the hyperfine dependence
of the atomic parity violation, was obtained only recently.
An important anapole moment bound in Tl also exists.
We discuss these measurements and their significance as tests
of the hadronic weak interaction, focusing on the mechanisms
that operate within the nucleus to generate the anapole moment.
The atomic results place new constraints on weak meson-nucleon
couplings, ones we compare to existing bounds from a variety
of $\vec{p}-p$ and nuclear tests of parity nonconservation.
\end{abstract}
\pagebreak

\maketitle

\section{INTRODUCTION}

Until 1957 physicists assumed that the fundamental laws of nature did not distinguish
between left and right.  However in that year,
following a suggestion by Lee and Yang~\cite{leeyang57}, experimenters discovered that
the weak force governing processes such as muon decay and nuclear beta decay violated
mirror symmetry maximally (to the accuracy of the measurements)~\cite{wu57}.  
Very soon afterwards Vaks and Zeldovich~\cite{zeldovich58} noted that weak interactions
would then modify the electromagnetic couplings to elementary particles (as well as
composite systems), allowing them to have a new, parity-violating moments, called 
anapole moments, in addition to their familiar parity-conserving ones (e.g.,
the charge and magnetic moments).  

This new moment has a number of curious properties.  It vanishes when probed by real
photons, and thus must be tested in processes where a virtual photon is exchanged.
Thus, for example, the anapole moment of a nucleus can be probed in electron
scattering and can influence the energies of bound electrons in an atom,
but cannot be measured through direct interactions with an electric field
(unlike magnetic moment interactions in a static $\vec{B}$ field).
The resulting electron-nucleus interaction in an atom is pointlike, thus
mimicking the short-range tree-level weak interaction induced by $Z^0$
exchange between atomic electrons and the nucleus.  The atomic cloud feels 
the nuclear anapole moment only to the extent that the orbiting electrons
penetrate the nucleus.  The anapole moment is an electric dipole coupling 
that is nuclear spin-dependent: it is this spin dependence, as we will see,
that allows anapole effects to be separated from tree-level weak
interactions.  Finally, the anapole moment is in general one of a larger
class of weak radiative corrections.  The anapole interaction -- one diagram is given
in Fig. 1a -- is thus accompanied by other radiative corrections
that do not correspond to virtual photon exchange, such as Fig. 1b.
It is the sum of all such diagrams that contribute to physical observables.
It follows that the anapole moment is not a measurable, that is, not 
separately a gauge-invariant quantity.  (However, we will later see that
the dominate contribution to nuclear anapole moments is well defined and
separately gauge invariant.)
  
\begin{figure}[!ht]
\psfig{bbllx=28pt,bblly=38pt,bburx=455pt,bbury=260pt,figure=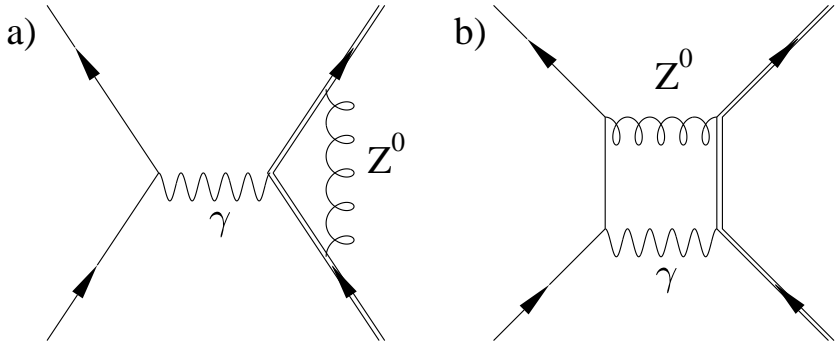,height=5.8cm}
\caption{Weak radiative corrections to electron-proton scattering
include anapole contributions (a) as well as conributions that
do not correspond to a single virtual photon exchange (b).}
\end{figure}
  
Our primary focus in this review is the nuclear anapole moment contribution to 
atomic parity nonconservation (PNC).  Exquisitely precise (sub 1\%)
measurements of atomic PNC have become possible in the past few 
years~\cite{bouchiat97}.  The primary focus of these studies has been to obtain
precise values of the strength of direct $Z^0$ exchange between
electrons and the nucleus.  The PNC effects are dominated by the
exchange involving an axial $Z^0$ coupling to the electrons and a
vector coupling to the nucleus.  The nuclear coupling is
thus coherent, proportional to the nuclear weak vector charge 
(approximately the neutron number), and independent of the
nuclear spin direction.  It is widely recognized that these
atomic measurements are important tests of the standard electroweak
model and its possible extensions, complementing what has been
learned at high energy accelerators that directly probe physics near the
$Z^0$ pole~\cite{erler,casalbuoni,rmusolf99}.  The comparison between precision measurements  
at atomic energies and accelerator energies could reveal the subtle
influence of new interactions beyond the standard model.

At low energies one expects weak radiative corrections, including 
the anapole contribution, to interfere with the dominant tree-level
weak amplitude, producing corrections to observables of relative
size $\alpha \sim$ 1\%, where $\alpha$ is the fine structure 
constant.  Therefore identifying the anapole contribution in weak
processes would seem a daunting task.  The possibility
that nuclear spin-dependent PNC measurements in heavy atoms might
prove an exception was first pointed out two decades ago
by Flambaum, Khriplovich, and collaborators~\cite{flambaum80}. 
Such spin-dependent PNC effects involve a vector coupling to the 
electrons and an axial coupling to the nucleus.  In this case the contribution 
of $Z^0$ exchange is considerably reduced because the 
nuclear interaction is no longer coherent, but instead 
samples, in a naive picture, only the spin of the last, unpaired 
nucleon within an odd-$A$ nucleus.
Additional suppression comes from the small
$Z^0$ vector coupling to the electron,
$\sim (4 \sin^2 \theta_W - 1)/2 \sim -0.05$.  In contrast, the
anapole moment is enhanced in heavy nuclei, growing
as $A^{2/3}$, where $A$ is the atomic number, and thus increasing in
proportion to the nuclear surface area.  The net result is the 
surprising conclusion that the anapole contribution -- the
weak radiative ``correction'' -- will in fact dominate over
the tree-level direct $Z^0$ spin-dependent contribution for 
nuclei heavier than $A$ $\sim$ 20.  

The principal motivation for this review was the recent successful
determination of the nuclear spin-dependent contribution to
atomic PNC in $^{133}$Cs~\cite{wood97}.  The extraction of this contribution from the much 
larger coherent PNC signal was accomplished by studying the 
dependence of the PNC signal on the choice of hyperfine level.
As the hyperfine differences are quite small, this extraction
requires detailed control of possible sources of systematic
error.  Over a decade ago the Colorado group~\cite{noecker88}
succeeded in measuring PNC effects in $^{133}$Cs to 2.2\%,
providing a tentative identification of the anapole moment.
Six years ago the Seattle~\cite{vetter95} and Oxford~\cite{edwards95}
groups achieved an accuracy of 1.2\% and 2.9\%, respectively, 
in PNC measurements in natural Tl, resulting in bounds
on the anapole moment in this nucleus.  Finally, in 1997 a definite measurement
was made: the Colorado group of Wood {\it et al.}~\cite{wood97} reported 
PNC measurements in $^{133}$Cs at the 0.35\% level, from which
a 7$\sigma$ nuclear spin-dependent PNC signal
was extracted.  This spin dependence is clearly well above
the expected signal from $Z^0$ exchange alone but compatible with
the enhancement expected from anapole effects.

As we will describe in this review, a
nucleus generates an anapole moment through the weak NN interaction,
which operates at long range in the nucleus by meson exchange,
where one meson-nucleon vertex is strong and the other weak~\cite{DDH,adelberger}.
Such interactions admix into the nuclear ground state odd-parity
amplitudes and also induce new PNC nuclear currents.  Thus
the exquisite precision now achieved in atomic PNC studies has
opened up a new window on the hadronic weak interaction.  This
interaction has proven more elusive than the weak interactions 
involving leptons.  Whereas the charge-changing hadronic weak interactions
can be studied in strangeness- or charm-changing decays, the 
standard model predicts that neutral-current interactions do not
change flavor.  Thus hadronic interactions mediated by the $Z^0$
can only be studied in NN interactions, where PNC must be exploited
to separate this contribution from much larger strong and 
electromagnetic effects.  This is a difficult task, and only a
few hadronic experiments have achieved the requisite precision and
even fewer (notably those done with $\vec{p}+p$ scattering or in
certain light nuclei with special properties) can be interpreted,
reasonably free of nuclear structure uncertainties.  The importance of
atomic anapole moment measurements is not only that they 
supplement the hadronic data, but also that they are sensitive
to long-range pion-exchange PNC interactions where the effects of
the neutral current can be isolated.  We will see, however, that
the extraction of weak coupling constant constraints from anapole
moments are requires some nonrivial nuclear physics analysis.

The plan of this review is as follows.  In Section 2 we 
discuss general properties of anapole moments, the 
associated current distribution, and the spin-dependent
interaction nuclear anapole moments generate in atoms.
In Sections 3 and 4 we summarize the experimental status
of atomic anapole moment measurements and the progress theory
has made in estimating the size of nuclear anapole moments from an 
underlying model of the hadronic weak interaction.  In 
Section 5 we compare weak meson-nucleon coupling constraints obtained from the
$^{133}$Cs anapole result and Tl limits  
with similar constraints obtained from 
PNC nuclear observables such as
$\vec{p} + p$ scattering and the circular polarization of 
gamma rays emitted from $^{18}$F.  It becomes clear that, 
while all measurements are consistent with the broad ``reasonable
ranges'' for PNC meson-nucleon couplings that theorists have
defined~\cite{DDH}, there is some level of disagreement between various
experiments when a global fit is performed.  We conclude 
by discussing prospects for improving our experimental
and theoretical knowledge of anapole moments.
  
\section{ANAPOLE MOMENTS AND ATOMS}

In this section we discuss general properties of nuclear anapole moments 
and the interactions they induce in atoms like $^{133}$Cs
and Tl.

\subsection{Electromagnetic Moments}

A standard multipole decomposition~\cite{walecka} of the electromagnetic 
current groups interactions according to their multipolarity and
symmetry properties into charge ($\hat{C}_J$), transverse electric ($\hat{T}_J^{el}$), and
transverse magnetic ($\hat{T}_J^{mag}$) operators, where $J$ denotes the multipole rank.
Static moments correspond to diagonal matrix elements of these
multipole operators.
From the transformation properties of the ordinary electromagnetic
current operator under
parity ($P$) and time-reversal ($T$) 
\begin{eqnarray}
\hat{P} J^\mu(\vec{x},t) \hat{P}^{-1} = J_\mu(\vec{x},t) \nonumber \\
\hat{T} J^\mu(\vec{x},t) \hat{T}^{-1} = J_\mu(\vec{x},-t),
\end{eqnarray}
and the constraint of hermiticity, it is readily verified that the
non-zero moments arise from matrix elements 
of the even-$J$ projections of $C_J$ and
the odd-$J$ projections of $T_J^{mag}$~\cite{walecka}.

More possibilities arise if one turns on the weak interaction, which 
introduces both parity- and CP/T-violating terms.  Thus it is
interesting to classify possible static moments more generally,
without the assumption that the underlying Hamiltonian conserves
electromagnetic symmetries.  The results in Table~\ref{table1}
show that, in addition to the ordinary monopole charge and
magnetic dipole moments, two new dipole moments then arise.  One
of these, corresponding to the expectation value of the $C_1$ multipole,
requires both parity and $CP/T$ violation.  This is the electric 
dipole moment which, while allowed in the standard model due to
$CP$ violation in the quark mass matrix and in the $\theta$ term,
has yet to be detected experimentally.  The second, an $E1$ 
moment, requires only parity violation.  This is the anapole
moment.

\begin{table}
\caption{Transformation properties of electromagnetic moments}
\label{table1}
\begin{center}
\begin{tabular}{cccc}
\hline \hline
$J$ & $C_J$ & $T^{el}_J$ & $T^{mag}_J$ \\
\hline
0 & $P$-even,~$T$-even & & \\
1 & $P$-odd,~$T$-odd & $P$-odd,~$T$-even & $P$-even,~$T$-even \\
2 & $P$-even,~$T$-even & $P$-even,~$T$-odd & $P$-odd,~$T$-odd \\
3 & $P$-odd,~$T$-odd & $P$-odd,~$T$-even & $P$-even,~$T$-odd \\
\hline
\end{tabular}
\end{center}
\end{table}

These monopole and dipole moments must therefore arise in the most
general expression for the matrix element of a conserved
four-current for a spin-${1 \over 2}$ particle
\begin{eqnarray}
\bar{U}(p') J^\mu(q) U(p) &=&
\bar{U}(p') (F_1(q^2) \gamma^\mu - i {F_2(q^2) \over 2 M} \sigma^{\mu \nu} q_\nu \nonumber \\
&& + {a(q^2) \over M^2} (\not q q^\mu - q^2 \gamma^\mu) \gamma_5
-i {d(q^2) \over M} \sigma^{\mu \nu} q_\nu \gamma_5 ) U(p).
\end{eqnarray}
The two vector terms define the charge $F_1(q^2)$ and 
magnetic $F_2(q^2)$ form factors.  The axial terms that
follow are the anapole and electric dipole terms, respectively.
The anapole term reduces
in the nonrelativistic limit to
\begin{eqnarray}
{a(q^2) \over M^2} (\not q q^\mu - q^2 \gamma^\mu) \gamma_5
&\rightarrow& {a(q^2) \over M^2} \vec{q}^2 (\vec{\sigma} - \hat{q} \hat{q}
\cdot \vec{\sigma}) \nonumber \\
&=& {a(q^2) \over M^2} \vec{q}^2 \vec{\sigma}_\perp,
\end{eqnarray}
showing that the current is transverse and spin dependent.
We define the anapole operator as
\begin{equation}
\hat{A}_{1 \lambda} = a(0) \sigma_{1 \lambda}.
\end{equation}

Weak radiative corrections that
modify physical processes involving elementary fermions
thus include anapole moment contributions.
One interesting example is the Majorana neutrino, where 
the identity under particle-antiparticle conjugation requires the
magnetic and electric dipole moments to vanish, but permits an
anapole moment~\cite{majorana}.  Yet the interest in anapole moments would likely
have remained largely theoretical were it not for the realization
that they might play a significant role in atomic PNC expriments.

\subsection{Atoms and the Generalized Siegert's Theorem}

A generalization of the anapole operator is helpful for the case
of a composite system like a nucleus, where the current 
operator and wave functions are modified by the interactions
among the constituents, inducing parity
admixtures in the ground state and multi-body
currents.  In the standard multipole decomposition the
$E1$ projection of the current is expressed in terms of the
multipole operator $\hat{T}^{el}_{1 \lambda}$
\begin{equation}
\vec{J}_{1 \lambda}(\vec{q}) \stackrel{E1}{\rightarrow} -i \sqrt{6 \pi}
\hat{T}^{el}_{1 \lambda}(q),
\end{equation}
where
\begin{equation}
\hat{T}^{el}_{1 \lambda}(q) = {1 \over q} \int d^3x \vec{J}(\vec{x}) \cdot
\vec{\nabla} \times (j_1(qx) \vec{Y}_{1 1 \lambda} (\Omega_x)).
\end{equation}

It is well known that current conservation places constraints on the
matrix elements of $\hat{T}^{el}_{J \lambda}(q)$.  A familiar example 
is the long-wavelength limit of $\hat{T}^{el}_{1 \lambda}(q)$ generated 
from the ordinary, parity-conserving electromagnetic current.
The operator then is $\vec{p}/M$, which is of order $v/c$, where $v$ is the nucleon velocity.  It 
can be shown that the exchange-current contributions to the 
vector three-current are also of order $v/c$.  For realistic models
of the nucleus which account for the interactions among the 
nucleons, in general we lack a prescription for constructing
interactions and currents consistently -- and for renormalizing 
them appropriately to take into account the limited Hilbert 
spaces employed in nuclear models.  As a result, there will be
errors in evaluations of $\hat{T}^{el}_{1 \lambda}$, owing to the
imperfect construction of the current, that are necessarily of
leading order in the velocity, $v/c$.  

Siegert~\cite{siegert} showed that the situation could be greatly improved by 
exploiting the continuity equation
\begin{equation}
\vec{\nabla} \cdot \vec{J}(\vec{x}) = -i [H,\rho(\vec{x})]
\end{equation}
to rewrite $\hat{T}^{el}_J$, in the long wavelength limit, entirely in terms 
of the charge operator.  This generates the familiar dipole form
of $\hat{T}^{el}_J$, proportional to $\omega \vec{r}$, where $\omega$ is the
energy transfer.  The importance of this rewriting is that the 
charge operator, which is of order $(v/c)^0$, has exchange current
corrections only of order $(v/c)^2$, or of relative size $\sim$ 1\%.
Thus the Siegert's form of $\hat{T}^{el}_J$, in which the constraints of
current conservation are fully exploited, is a far more controlled
operator for use in nuclear calculations.  

The analogous situation arises for the anapole moment.  While 
there are a variety of forms of the anapole operator that are equivalent
up to terms that vanish by current conservation, these forms
are not equivalent operationally in realistic calculations 
because of model violations of current conservation.  The simple,
long-wavelength form of Siegert's theorm
doesn't address the question of moments because it generates an
operator proportional to $\omega$, which then
vanishes for a diagonal matrix element.  Fortunately the
generalization of Siegert's theorem~\cite{friar} exists: at arbitrary $q$ one
can write $\hat{T}^{el}_J(q) = \hat{S}_J(q) + \hat{R}_J(q)$, where all components
of the electromagnetic current that are constrained by current conservation
have been isolated in $\hat{S}_J$ and expressed as a commutator
of the charge operator with the nuclear Hamiltonian.  All such
terms then vanish for a static moment.  The resulting generalized
Siegert's form of $\hat{T}^{el}_J$ appropriate for diagonal matrix elements
is
\begin{equation}
\left. \hat{T}^{el}_{1 \lambda}(q) \right|_{diagonal} \buildrel \vec{q} \rightarrow 0 \over
\rightarrow - {i \vec{q}^2 \over 9 \sqrt{6 \pi}} \int d^3r r^2
[J_{1 \lambda}(\vec{r}) + \sqrt{2 \pi} (Y_2(\Omega_r) \otimes
J_1(\vec{r}))_{1 \lambda}].
\end{equation}

From Eqs. (5) and (8) we have the appropriate threshold form of the
current operator.  The anapole operator is defined relative to the
current operator as in Eqs. (3) and (4), yielding immediately~\cite{haxton89}
\begin{equation}
\hat{A}_{1 \lambda} = - {M^2 \over 9} \int d^3r r^2
[J_{1 \lambda}(\vec{r}) + \sqrt{2 \pi} (Y_2(\Omega_r) \otimes
J_{1}(\vec{r}))_{1 \lambda}]
\end{equation}
Other forms of the anapole operator are more commonly used, for
example~\cite{khriplovich}
\begin{equation}
\hat{A}_{1 \lambda} = - \pi \int d^3r r^2 J_{1 \lambda}(\vec{r}).
\end{equation}
Apart from the normalization, which is a matter of convention, this
form will be equivalent to Eq. (9) only if the nuclear model is
sufficiently simple that exact current operators can be 
constructed.  This would be the case for a simple central potential
model, such as an independent-particle harmonic oscillator, for
which the appropriate current operator is that of a free particle.
This would also be the case for an independent-particle model with
spin-orbit and orbital potentials of the form $\vec{\sigma} \cdot \vec{\ell}$
or $\vec{\ell} \cdot \vec{\ell}$, provided that the additional contributions
to the current obtained by minimal substitution, $\vec{p} \rightarrow
\vec{p} - e \vec{A}$ into $\vec{\ell} = \vec{r} \times \vec{p}$,
are included in the calculation~\cite{musolfphd}.  But in other, more realistic
treatments of the nuclear physics, Eq. (9) is the unique form that fully enforces the
constraints of current conservation, regardless of the complexity
of the current operators.

Figure 2 gives a classical picture of the anapole moment as a 
current winding within the nucleus.  Although the currents
on the inner and outer sides of the torus oppose one another,
there is a net contribution to Eq.(9) because of $r^2$ weighting
of the current, leading to an anapole moment that points
upward.  (Note the sign in Eq. (9).)  The current distribution drawn
in Fig. 2 is odd under reversal of parity, as is the ordinary
$J_{1 \lambda}^{em}$.  Thus it is easy to see that the 
corresponding anapole moment is a parity-odd operator.  If,
however, the current has a chirality -- a small ``pitch''
corresponding to a left- or right-handed winding that 
would signal PNC -- a parity-even contribution to the 
operator would be induced.  This is the analog of evaluating
Eq. (9) with the axial-vector current induced by the weak
interaction, leading to an operator component that would 
have a nonzero expectation value for a parity-conserving
ground state.  Similarly, the anapole moment associated with
$J^{em}_{1 \lambda}$ would have a nonzero expectation value
if the nuclear ground state contains parity-odd components
induced by the weak PNC component of the $NN$ interaction.

\begin{figure}[!ht]
\psfig{bbllx=-20pt,bblly=0pt,bburx=360pt,bbury=260pt,figure=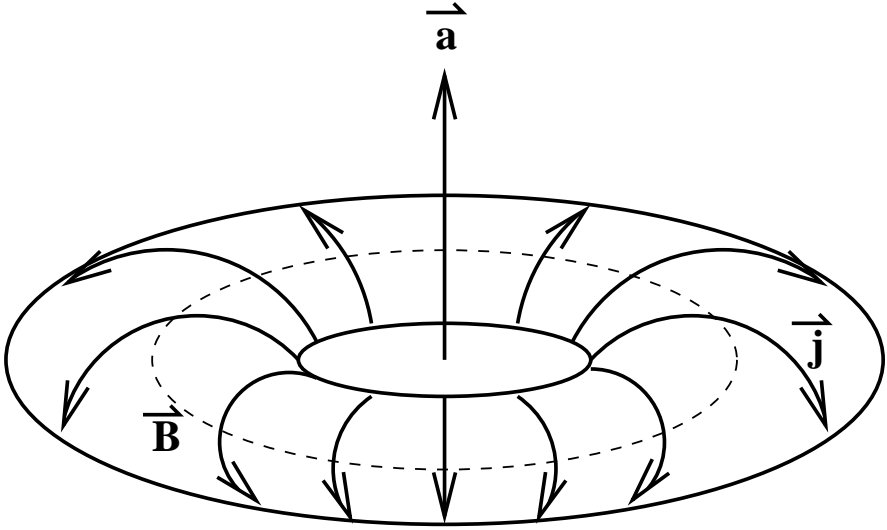,height=7.0cm}
\caption{A toroidal current winding that would correspond to 
a nonzero anapole operator in Eq. (9).}
\end{figure}
  
\subsection{The Spin-Dependent Atomic Interaction}

The potential felt by atomic electrons due to the nuclear anapole moment is generated in
the standard way by the dipole current - dipole current interaction.
As the $\vec{q}^2$ appearing in the nuclear current cancels the 
photon propagator for a static interaction, this weak interaction
has the contact form
\begin{equation}
H_W^{s.d.} = {G_F \over \sqrt{2}} \kappa \vec{\alpha} \cdot \vec{I} \rho(\vec{r}),
\end{equation}
where $\vec{I}$ and $\rho(\vec{r})$ are the nuclear spin and density,
and where $\vec{\alpha}$ is the Dirac matrix acting on the 
electrons.  The superscript $s.d.$ denotes that this is the nuclear
spin-dependent contribution to the atomic PNC interaction.
The portion of this interaction generated by the
nuclear anapole moment is then 
\begin{equation}
\kappa_{anapole} = {4 \pi \alpha \sqrt{2} \over M^2 G_F}
{<I || \hat{A}_1 || I>/e \over <I||\hat{I}||I>}
\end{equation}
where $||$ denotes a matrix element reduced in angular momentum
and $\alpha$ is the fine structure constant.
The reduced matrix element of $\hat{I}$ is $\sqrt{I(I+1)(2I+1)}$.
[Note that elsewhere in the literature another definition of $\kappa$
is more commonly used, one that associates the nuclear
ground state with a single-particle level of orbital
angular momentum $\ell$ and spin $I$.  For example, the $\kappa$
used in Ref.~\cite{flambaum80} is obtained by dividing ours
by the factor $(-1)^{I+1/2+\ell} (I+1/2)/(I(I+1))$.]
  
We have already noted that $\kappa_{anapole}$ is just one of the
contributions to $\kappa$: isolating this anapole contribution would
be clearly easier if $\kappa_{anapole}$ were the dominant such
contribution.  It was the important observation of Flambaum and
Khriplovich~\cite{flambaum80} that the $A^{2/3}$ growth of nuclear anapole moments,
already implicit in the $r^2$ weighting of the current in 
Eq. (9), would lead to such dominance in heavy atoms.  It is also
clear that, if $\kappa_{anapole}$ could be extracted from atomic
measurements, relating this quantity to the underlying 
sources of hadronic PNC involves specifying the PNC currents 
that contribute to Eq. (9), as well as the PNC admixtures in the
nuclear ground state wave function through which the ordinary
electromagnetic current operator will generate nonvanishing matrix
elements of $\hat{A}_{1 \lambda}$.  We will defer this question to
Section 4, focusing now on the experimental progress that has
produced values and limits for $\kappa$.

\section{EXPERIMENTS}

The idea of the anapole moment languished for many 
years after the early work of Zeldovich and Vaks. 
In 1973 Henley, Huffman, and Yu~\cite{henley} -- who were unaware of the Zeldovich
paper -- noted that the anapole moment would contribute to PNC 
observables in high energy electron scattering off the proton.
Then a revival of interest in the anapole moment occurred in 1980 when
its possible relevance to ongoing experiments
on atomic PNC was noticed~\cite{flambaum80}.  In these
experiments the tiny parity-violating mixing of $S$ and $P$ states  
induced by the coherent, nuclear-spin-independent $Z^0$ coupling
to the nucleus was being measured~\cite{bouchiat74}.  As both the coherence and
the increased electronic overlap with the nucleus lead
to larger PNC effects in heavy atoms,
the experiments focused on atoms with $A \gsim 100$.  The $A^{2/3}$
enhancement of the anapole moment
makes the spin-dependent contribution to atomic PNC
dominant over other 
vector(electron)-axial(nucleus) sources of PNC in such heavy atoms.
Spin-dependent effects of the expected size -- a few percent of the 
spin-independent signal -- could, in principle, be extracted
from the experiments by precisely comparing the amount of mixing for two
different hyperfine components, {\it i.e.,} electronic transitions that 
differ only in the orientation of the nuclear spin.   

Nevertheless, the spin-dependent PNC effects are painfully small, corresponding
to state mixings on the order of parts in $10^{13}$.  That level of 
precision was eventually reached in a cesium PNC experiment 
and nearly reached in experiments with
thallium.  The isolation of a definite spin-dependent contribution
in cesium provided the first confirmation
that anapole moments exist.

\subsection{The Cs Experiment: Techniques}

The experiment in cesium is described in detail in Refs.~\cite{wieman1}. 
The central idea is to exploit the highly forbidden
6$S$-to-7$S$ transition in atomic cesium.  The strength of the 
transition depends on the handedness within the excitation region.  
That handedness is defined by various applied electric, magnetic, 
and laser fields and reverses with appropriate reversals of those 
applied fields.  This reversal forces any PNC 
component to the excitation rate to change sign, thereby 
altering the excitation rate for the transition.   

\begin{figure}[!ht]
\psfig{bbllx=30pt,bblly=56pt,bburx=510pt,bbury=450pt,figure=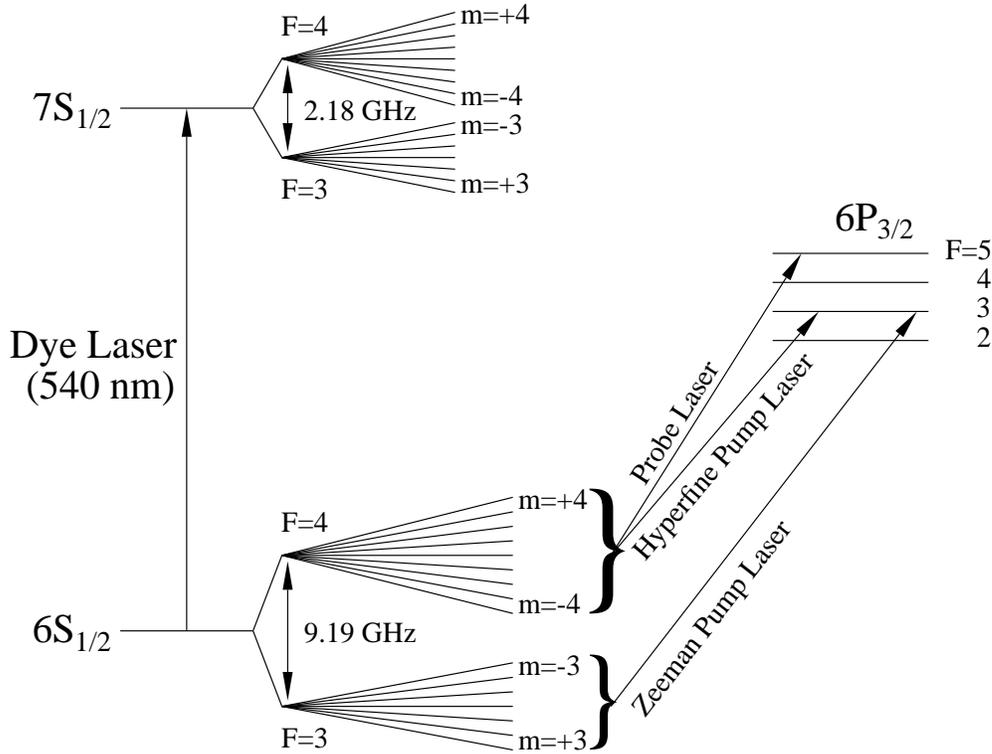,height=10.7cm}
\caption{Partial cesium energy-level diagram including the 
splitting of the $S$ states by the magnetic field.  The case of
540-nm light exciting the $(F=3,m_F=3)$ level is shown.  Diode
lasers 1 and 2 optically pump all of the atoms into the (3,3)
level, and laser 3 drives the 6$S_{F=4}$ $(F_{det})$ to
$6P_{F=5}$ transition to detect the $7S$ excitation.  PNC is
also measured for excitation from the (3,-3), (4,4), and (4,-4)
$6S$ levels.  The diode lasers excite different transitions
for the latter two cases.}
\end{figure}
  
As the fractional change in the excitation rate is very small,
a great deal of work must be done to achieve a signal-to-noise
ratio sufficient to see any effect.  Then even more work must be invested 
to verify that the detected effect is truly a 
violation of parity and not a spurious signal arising from 
systematic errors such as imperfect reversals or alignments of the 
fields that define the handedness of the experiment. 

In the absence of electric fields or parity violating 
interactions, the electric dipole ($E1$) transition between the $6S$ 
and $7S$ states of the cesium atom (Fig. 3) is forbidden.
As the nuclear spin of $^{133}$Cs is $I$ = 7/2,
these $S_{1/2}$ levels combine with the nuclear ground state to
form hyperfine states of total angular momentum $F$ = 4 and 3.
A PNC interaction mixes a small
amount ($\sim 10^{-11}$) of the neighboring $6P_{3/2}$ and $7P_{3/2}$ states into the $6S$ and $7S$ states: 
the $P_{3/2}$ hyperfine levels have $F$ = 2-5, so that PNC mixing
with both the $F$ = 3 and 4 hyperfine $S$ states takes place.
The induced $7S \leftrightarrow 6S$ PNC $E1$ amplitude is thus 
proportional to the product of the PNC mixing and $E1$ matrix
elements coupling the $S$ and $P$ hyperfine levels (and 
inversely proportional to the energy differences between the
$S$ and $P$ levels).
To obtain a measurable observable that is first order in this small 
amplitude, a d.c. electric field $\vec{E}$ is applied that also mixes $S$ and $P$ 
states.  This field generates an interfering ``Stark-induced'' $E1$ transition
amplitude $A_E$ that is typically $10^5$ times larger than  
the PNC-induced $E1$ amplitude $A_{PNC}$.
A complete analysis of the relevant transition 
rates between the various hyperfine magnetic states
$(F,m_F)$ is given in Ref.~\cite{wieman1}.
These rates involve a straightfoward evaluation of the hyperfine
matrix elements for the Stark amplitude, the nuclear-spin-dependent
PNC Hamiltonian given in Eq. (11), and the  
spin-independent PNC Hamiltonian for the much stronger coherent interaction with
the nucleus,
\begin{equation}
H_W^{s.i.} = {G_F \over 2 \sqrt{2}} Q_W \gamma_5 \rho(\vec{r}).
\end{equation}
Here $\gamma_5$ is the axial coupling to the electrons, while the
tree-level standard-model result for the vector weak charge of a point
nucleus is 
\begin{equation}
Q_W = Z(1-4 \sin^2 \theta_W) - N \sim -N.
\end{equation}
We omit the somewhat tedious angular momentum algebra here.
To generate a nonzero interference between $A_{PNC}$ and $A_E$,
which differ by a relative phase of $i$, the $6S$-to-$7S$
transition must be excited with an elliptically polarized laser field 
of the form $\epsilon_z \hat{z}$ + $p$ $i$ Im($\epsilon_x$) $\hat{x}$, where the handedness of the 
polarization $p$ is $\pm$ 1, $\epsilon_x$ is the component of the
oscillating electric field parallel to the d.c. Stark field $\vec{E}$,
and $\epsilon_z$ is the oscillating field component in the direction
perpendicular to $\vec{E}$.  The PNC contribution to the 
transition rate is measured on both the $6S_{F=3}$-to-$7S_{F=4}$ and 
$6S_{F=4}$-to-$7S_{F=3}$ transitions.  To a good 
approximation the only difference between the two transitions is 
the reversal of the nuclear spin.  Thus simply taking the 
difference between the PNC contributions to the 
hyperfine transition rates isolates the nuclear-spin-dependent PNC 
coupling $\kappa$ of Eq. (11).

For both transitions the atoms are initially populated and 
excited only out of states with extreme values of $m_F$,
$(F = 3, m_F = \pm 3)$ and $(F = 4, m_F = \pm 4)$.  For
these cases and the configuration of electric and magnetic fields 
shown in Fig. 4, the transition rate is
\begin{eqnarray}
R &=& |A_E + A_{PNC}|^2 \sim \beta^2 E_x^2 \epsilon_x^2 C_1(F,m_F;F',m_F') \nonumber \\
&+& 4 \beta E_x \epsilon_x p \mathrm{Im}(\epsilon_x) \mathrm{Im}(E1_{PNC}) C_2(F,m_F;F',m_F')
\end{eqnarray}
where $\beta$ is the tensor transition polarizability, $E_x = |\vec{E}|$ is the d.c. 
electric field strength, and $C_1$ and $C_2$ are combinations of Clebsch-Gordan
coefficients whose detailed form we have suppressed~\cite{wieman1}.
They depend on the initial- and final-state hyperfine labels $(F,m_F)$ and
$(F',m_F')$ and transform as $C_1(m_F) \rightarrow C_1(-m_F)$
and $C_2(m_F) \rightarrow -C_2(-m_F)$ under reversal of 
the magnetic labels.  The tiny contribution proportional to
$A_{PNC}^2$ has been neglected as
well as the small $6S-7S$ parity-conserving magnetic dipole transition amplitude 
$A_{M1}$. (As discussed later, $A_{M1}$ can be a source of many 
systematic errors that must be addressed carefully.)
  
\begin{figure}[!ht]
\psfig{bbllx=17pt,bblly=206pt,bburx=484pt,bbury=512pt,figure=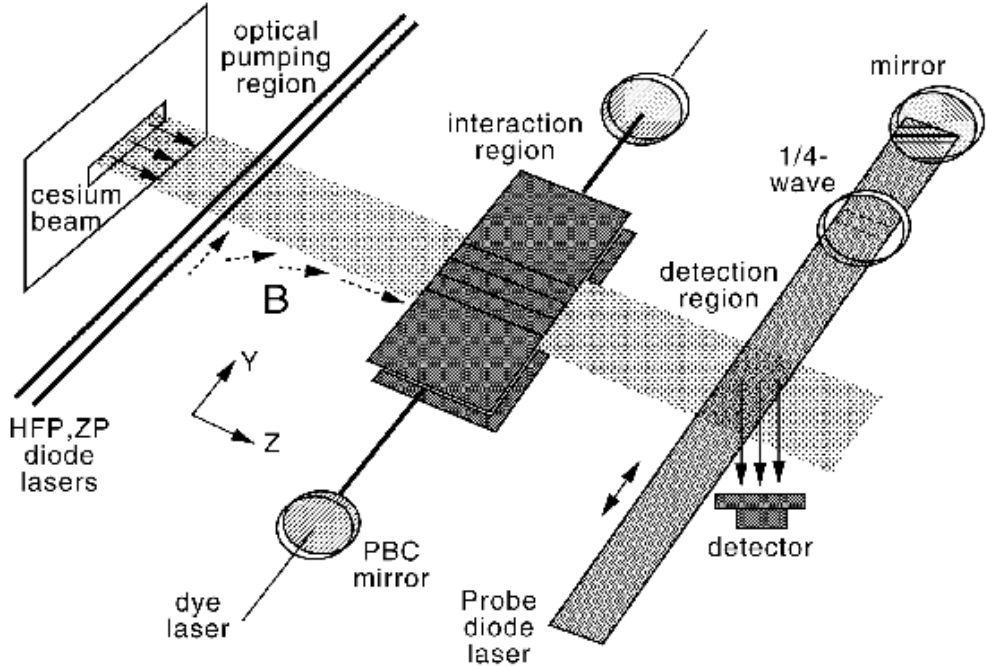,height=9.0cm}
\caption{Schematic of the cesium apparatus.  In the interaction 
region $\vec{B}$ is along the $\hat{z}$ axis, $\vec{E}$ is along
the $\hat{x}$ axis, and the 540-nm dye laser beam defines the
$\hat{y}$ axis.}
\end{figure}
  
The experimental quantity of interest is the fractional 
PNC modulation in the transition rate  
\begin{equation}
{\Delta R \over R} = {2 \mathrm{Im}(\epsilon_x) \mathrm{Im}(E1_{PNC}) \over \epsilon_z \beta E}
\end{equation}
that modulates with the reversals of $E$, $m$, and $p$.
In the experiment, there are actually five ``parity'' 
reversals because $m_F$ is reversed in three different ways. The use 
of this large number (five) of independent reversals is essential 
for the detection and elimination of systematic errors, providing 
a great deal of redundancy for the 
PNC signal.  This redundancy means that although no single 
reversal is perfect, the product of all the imperfections is far 
smaller than the PNC signal.  Furthermore, the signal 
modulations accompanying various combinations of 
field reversals provide additional information about the field 
reversal imperfections and orientations that help identify 
potential systematic errors. 

\subsection{The Cs Experiment: The Apparatus}
  
A simplified schematic of the apparatus used by Wood {\it et al.}~\cite{wood97} is shown in 
Fig. 4.  An effusive beam of atomic cesium is produced by a 
heated oven and collimated with a multichannel capillary array nozzle. 
The beam is optically pumped into the desired $(F,m_F)$ 
state by light from diode lasers 1 and 2~\cite{wieman1,wieman2}.  The 2-
cm-wide beam of polarized atoms intersects the 540-nm 
standing wave (Gaussian diameter, 0.8 mm) that is inside a high-finesse
(100,000) Fabry-Perot power buildup cavity (PBC). 
The PBC not only enhances the transition rate, but the standing 
wave geometry also greatly suppresses the troublesome 
modulation arising from $A_{E1}-A_{M1}$ interference~\cite{wieman1}.  The 540-nm 
light originates from a dye laser whose frequency is tightly 
locked to the resonant frequency of the PBC by a high-speed 
servosystem~\cite{wieman3}.  Before entering the PBC, the dye laser light 
passes through an intensity stabilizer, an optical isolator, and a 
polarization control system made up of a half-wave plate, 
Pockels cell, and adjustable birefringence compensator plate. 
The ellipticity of the light is controlled by rotating the half-wave 
plate, and the handedness of the ellipse is reversed by switching 
the sign of the l/4 (quarter wavelength) voltage that is applied to 
the Pockels cell.  The intensity stabilizer holds the amount of light 
transmitted through the cavity constant and hence stabilizes the 
field inside of the cavity.  This field corresponds to about 2.5 kW 
of circulating power.  The PBC resonant frequency is held at the 
frequency of the desired atomic transition by a servosystem that 
translates the input mirror.

The d.c. electric field in the interaction region is produced 
by applying $\sim$ 500 volts between two parallel 5-cm by 9-
cm conducting plates, separated by about 1 cm.  The plates are 
made of flat pieces of Pyrex glass coated with 100 nm of 
molybdenum.  Both field plates were divided into five electrically 
separate segments.  This division makes it possible to apply small 
uniform and gradient electric fields along the y-axis for auxiliary 
diagnostic experiments.  The entire buildup cavity and field-plate 
mounting system is rather elaborate to ensure precise alignment 
and extreme mechanical stability.

After being excited out of the populated $6S$ hyperfine level up to 
the $7S$ state, an atom will decay by way of the $6P$ states to the 
previously empty $6S$ hyperfine level ($F_{det}$) more than 60\% of the time.  
The population of $F_{det}$ is detected 10 cm downstream of the 
interaction region using laser florescence.  Light from diode laser 
3 excites each atom in $F_{det}$ to the $6P_{3/2}$ state many times.  The 
resulting scattered photons are detected by a silicon photodiode 
that sits just below the atomic beam.  When the $6S_{F=3}$-to-$7S_{F=4}$ 
line is measured, the detection laser drives the $6S_{F=4}$-to-$6P_{3/2,F=5}$ 
cycling transition (Fig. 3).  About 240 photons per $F_{det}$ atom are 
detected. For the $6S_{F=4}$-$7S_{F=3}$ line, the detection transition is the 
$6S_{F=3}$-to-$6P_{3/2,F=2}$ cycling transition.  This cycle gives about 100 
detected photons per $F_{det}$ atom.  The signal-to-noise ratio for this 
transition is about 20\% lower.  During each half cycle of the most 
rapid field reversal ($\vec{E}$) and after the switching transient has 
passed, the detector photocurrent is integrated, digitized, and 
stored.  For each stored value, the computer also records the 
field and spin orientations.

The signal-to-noise ratio needed for this experiment puts 
extreme requirements on laser stability.  A fluctuation in the 
intensity, frequency, or direction of the light from any of the four 
lasers will introduce noise in the detected atomic fluorescence. 
The most extensive control is needed for the dye laser~\cite{wieman3}, 
but the requirements on the three diode lasers used for optical 
pumping and detection are also severe.  These requirements  
motivated substantial development of diode-laser stabilization 
technology~\cite{wieman4}.  Both optical and electronic 
feedback were used to lock the frequency of each diode laser to 
the desired atomic transition using saturated absorption 
spectrometers.

Extensive and precise control of magnetic fields was required in 
the experiment.  In the optical pumping region, there is a uniform 
2.5-G field that must point in the $\hat{y}$ direction (parallel to the 
pumping laser beams). In the interaction region, a 6.4-G field 
must point precisely along either $\pm \hat{z}$.  Between
the two regions, the magnetic field must rotate gently enough that 
the atomic spins follow it adiabatically.  Finally, the field must be 
near zero in the detection region, and it is necessary to precisely 
reverse the fields in the optical pumping and interaction regions 
independently without significantly perturbing the fields in the 
other two regions.  The setup required the use of 23 
magnetic field coils of various shapes to provide the necessary 
fields and gradients.  Most of these coils were driven with both 
reversing and nonreversing components of current.

Many additional elements were required to achieve sufficiently 
precise alignment and control of all aspects of the apparatus. 
These include 31 different servosystems to ensure optical, 
mechanical, and thermal stability.

\subsection{The Cs Experiment: Data and Results}
  
Approximately 350 hours of data on the PNC observables
were acquired in five runs distributed over an 8-month period. 
Each of the five runs followed the same basic procedure.  First, a 
set of auxiliary experiments was carried out 
to measure and set numerous quantities: (i) all three 
components of the average $\vec{E}$ and $\vec{B}$ fields and their $\hat{y}$ gradients 
in the interaction region; (ii) the magnitude and orientation of the 
birefringence of the PBC output-mirror coating; (iii) the 
polarization-dependent power modulation of the green laser 
light; and (iv) the populations of the $m_F$ levels of the atomic beam 
as it entered the interaction region.  After these measurements 
were completed, the four laser frequencies were locked to the 
desired hyperfine transitions and data were acquired in blocks of 
about 1.5 hours each.  During this time, five parity reversals -- 
the electric field, laser polarization, and three ways of reversing the 
$m_F$ state being excited (reversing the polarization of the optical 
pumping light of laser 2, reversing the optical pumping magnetic 
field relative to the pumping light, and reversing the magnetic 
field in the 6S-7S excitation region) -- were carried out at different rates. 
The electric field was
reversed at 27 Hz and the others at various lesser rates.  The 
relative phases of the various reversals were regularly shifted by 
one half cycle.  Before and after each of the 1.5-hour blocks, the 
polarization of the 540-nm standing-wave field was measured 
and set.  At regular intervals, the PBC output mirror was rotated 
by 90.0 degrees, and at irregular intervals, the frequencies of 
the four lasers were changed to measure PNC on the other $6S$-
$7S$ hyperfine line.  At the end of each data run (20 to 30 1.5-hour 
blocks of data), the initial auxiliary experiments were repeated to 
check that the quantities described above had not changed 
significantly.

The typical size of the $6S-7S$ signal from the photodiode 
was 200 nA on the 3-4 line and 85 nA on the 4-3 line. These 
measurements corresponded to about 0.5\% of the atomic beam 
undergoing a $6S-7S$ transition.  The signal-to-noise ratio for 
measuring the $6S-7S$ transition rate was typically 55,000/$\sqrt{Hz}$, 
and the parity violating modulation was typically about six parts 
in $10^6$.  Because the technical noise was small and many photons 
were detected from each atom that had undergone a $6S-7S$ 
excitation, the noise was dominated by the shot-noise 
fluctuations in the number of atoms making the $6S-7S$ transition. 
Data was taken over a range of polarization ellipticities and 
electric fields.  

The data analysis used to find the PNC modulation for 
each block of data was relatively simple.  Appropriate 
combinations of fractional differences in the signal sizes were 
calculated for each of the 32 different field configurations to find 
the fraction of the rate that modulated with all five reversals. 
From the measured values of electric field and laser polarization 
this fractional modulation was then converted into Im$(E1_{PNC})/\beta$. 
Various other small calibration corrections were required.  

The major concern in this experiment was possible 
systematic errors arising from spurious signals that modulate 
under all five parity reversals, thus mimicking PNC.  Roughly 20 
times more data were taken in the investigation and elimination 
of such errors than in the actual PNC measurement.  Several 
small errors associated with stray and misaligned fields were 
encountered, as in previous PNC measurements~\cite{wieman1}, and were 
treated as before.  An exhaustive analysis was carried out of 
all possible combinations of static and oscillating electric and 
magnetic fields that could mimic a PNC signal.  All of the stray 
(defined as nonreversing) and misaligned d.c. electric and 
magnetic fields and their gradients, and many of the laser field 
components, could be determined by looking at appropriate 
modulations in the $6S-7S$ rate under various conditions.  Many of 
these quantities were extracted from the 31 different modulation 
combinations observed in real time while taking PNC data, and 
the remaining components were determined by the auxiliary 
experiments that were interspersed with the PNC runs.  Many 
tests were also performed to ensure the necessary stability of the 
relevant fields. 

Although the procedures were similar in concept to 
previous work, the Wood {et al.}~\cite{wood97} measurements
were more difficult and time consuming
due to the higher accuracy required.  It proved necessary 
to consider not only the average fields, but also their 
gradients across the interaction region.   
Several small spurious signals were identified and removed. 
The absence of any systematic effects from the
troublesome $A_E - A_{M1}$ interference was confirmed 
by the independence of results for various laser polarizations
sensitive to this interference.
Other cross-checks included enhancing sources of error
to confirm the predicted response and performing data analyses
to verify that transition rate variations were consistent with
fundamental shot-noise fluctuations.

The final results are
\begin{eqnarray}
-\mathrm{Im}(E1_{PNC})/\beta &=& 1.6349 \pm 0.0080~\mathrm{mV/cm}~~6S_{F=4} \leftrightarrow
7S_{F=3} \nonumber \\
&=& 1.5576 \pm 0.0077~\mathrm{mV/cm}~~6S_{F=3} \leftrightarrow 7S_{F=4}
\end{eqnarray}
yielding for the nuclear-spin-dependent difference of interest 
for the anapole moment
\begin{equation}
\mathrm{Im}(E1_{PNC}^{s.d.})/\beta = 0.077 \pm 0.011~\mathrm{mV/cm},
\end{equation}
a $7\sigma$ effect.  The statistical uncertainties for the two
transitions, 0.0078 and 0.0073 mV/cm, respectively, dominate 
the error. The systematic uncertainties are based on statistical 
uncertainties in the determination of various calibration factors 
and systematic shifts, and therefore, it is appropriate to add them 
in quadrature. The final results are in good agreement with 
previous measurements in cesium but are much more precise.
From the best available atomic calculation~\cite{flambaum97,dzuba87}
one then finds
\begin{equation}
\kappa(^{133}\mathrm{Cs}) = 0.112 \pm 0.016,
\end{equation}
a result we will find is dominated by the anapole moment.

\subsection{The Tl Experiment}

Experiments measuring parity violation in atomic thallium
(70.5\% $^{205}$Tl, 29.5\% $^{203}$Tl)
have not yet detected a nuclear spin dependent/anapole moment 
contribution, but they have achieved accuracies very near the 
level where anapole effects are expected.  We will see that the
resulting limits are interesting from the perspective of
hadronic PNC.  An interference 
of parity-allowed and PNC contributions to an atomic transition
is observed, just as in cesium.  
However, in thallium the transition is an allowed 
magnetic dipole transition, $6P_{1/2} \leftrightarrow 6P_{3/2}$.  The PNC
interference is between the allowed M1 transition amplitude and 
the PNC E1 amplitude arising from weak interaction effects 
that mix $S$ states into the $P_{1/2}$ ground state.  As Tl
has an $I = 1/2$ nuclear ground state, the $P_{1/2}$ state
has $F=0$ and $F=1$ hyperfine sublevels which will be affected
differently by nuclear-spin-dependent sources of PNC.
Thus the search for spin-dependent PNC effects requires a comparison
of the strength of PNC transitions out of the $F=0$ and $F=1$ 
states.  Groups at the University of Washington~\cite{vetter95}
and Oxford~\cite{edwards95} have made such comparisons.
  
\begin{figure}[!ht]
\psfig{bbllx=20pt,bblly=28pt,bburx=540pt,bbury=290pt,figure=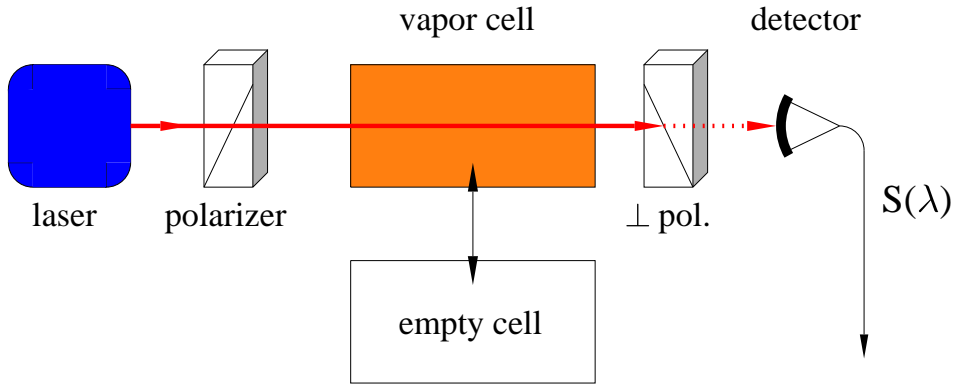,height=6.0cm}
\caption{Schematic of the thallium vapor cell apparatus.}
\end{figure}

The optical rotation of linearly 
polarized light that is produced by the interference of $E1$ and $M1$ 
transitions is measured in a large vapor cell of 
atomic thallium, as shown schematically in Fig. 5.  A beam of 
linearly polarized laser light passes through the vapor and then 
through a nearly crossed linear polarizer followed by a 
sensitive detector.  A Faraday modulator rotates the plane of 
polarization back and forth through the perfectly crossed 
position.  This allows phase-sensitive detection that minimizes 
sensitivity to drifts in the signal baseline.  The laser is scanned 
repeatedly across the transition and the detected light signal is 
then fit to the predicted lineshape.  The PNC rotation reverses 
sign as the laser is tuned through line center.  From  
the fit to the dispersion-like line shape the amplitude of the
rotation can be determined.  To distinguish the signal of
interest from spurious frequency-dependent rotations in optics, 
the vapor cell is interchanged with a ``dummy'' cell that has 
identical optics but no atomic vapor.   An example of the data 
obtained in this fashion is shown in Fig. 6, taken from Ref~\cite{vetter95}. 
  
\begin{figure}[!ht]
\psfig{bbllx=0pt,bblly=120pt,bburx=540pt,bbury=600pt,figure=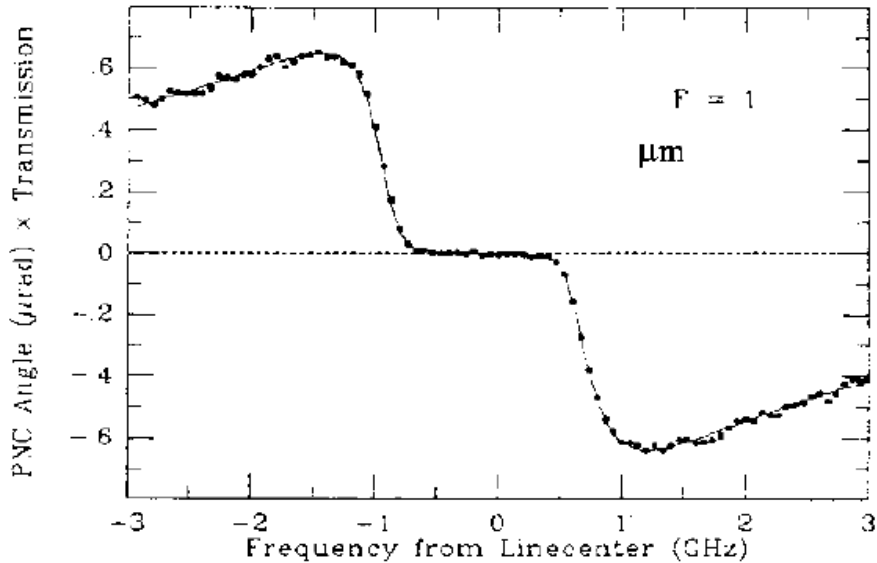,height=10cm}
\caption{Combined PNC optical rotation for the thallium data
cycles (a total of $\sim$ 230 h) for $F=1$ (points) compared
to the theoretical (dispersion $\times$ transmission) line
shape (solid line).}
\end{figure}
  
It can be seen that the agreement between predicted and 
observed signals is excellent.  As in the cesium experiment 
hundreds of hours of data were taken and considerable effort 
was made to eliminate possible systematic errors.  The thallium 
oven was carefully constructed and shielded to avoid magnetic 
fields that would cause spurious Faraday rotation.  Data were 
acquired over a wide range of pressures of thallium and a 
variety of tests of the calibration were carried out.  

The final results of the experiments are expressed as the 
ratio of parity violating $E1$ to magnetic dipole $M1$ transition 
amplitudes, $R = \mathrm{Im}(E1_{PNC}/M1)$.  The
nuclear-spin-independent PNC signals obtained are $R^{s.i.} =  
(-14.68 \pm 0.17) \times 10^{-8}$~\cite{vetter95} (normalized to $^{205}$Tl) 
and $(-15.68 \pm 0.45) \times 10^{-8}$~\cite{edwards95}, corresponding
to 1.2\% and 2.9\% accuracy.  The 
nuclear-spin-dependent effects are consistent with zero,
\begin{eqnarray}
R^{s.d.}(\mathrm{Tl}) &=& (0.15 \pm 0.20) \times 10^{-8}~~\mathrm{Seattle} \nonumber \\
 &=& (-0.04 \pm 0.20) \times 10^{-8}~~\mathrm{Oxford.} 
\end{eqnarray}
The resulting constraints on $\kappa$ are
\begin{eqnarray}
\kappa(\mathrm{Tl}) &=& 0.29 \pm 0.40~~\mathrm{Seattle} \nonumber \\ 
&=& -0.08 \pm 0.40~~\mathrm{Oxford}.
\end{eqnarray}

The $^{133}$Cs and Tl spin-dependent measurements 
involve odd-proton nuclei, which leads to the prediction that they 
test a similar combination of isospin components of the weak 
hadronic PNC potential, as we will see below.  Note that
an anapole measurement for an odd-neutron nucleus 
would test a roughly orthogonal combination of isoscalar
and isovector contributions to the hadronic PNC potential.
Several atomic PNC efforts are
underway that could produce such constraints~\cite{budker,fortson}.

\section{ANAPOLE MOMENTS AND HADRONIC PNC}

In this section we discuss the extraction of $\kappa_{anapole}$ from $\kappa$, 
the relation between $\kappa_{anapole}$ and the weak hadronic 
interaction, and the mechanisms by which a nucleus generates an
anapole moment.

\subsection{Extracting the Anapole Moment}

We have seen that atomic PNC measurements place the following
constraints on the strength of nuclear-spin-dependent 
electron-nucleus interaction
\begin{eqnarray}
\kappa(^{133}\mathrm{Cs}) &=& 0.112 \pm 0.016 \nonumber \\
\kappa(\mathrm{Tl}) &=& 0.29 \pm 0.40.
\end{eqnarray}
where we have used the Seattle Tl result above because, due to its
central value, it proves to be the more restrictive in the 
weak meson-nucleon parameter region favored by nuclear PNC experiments.
The three principal contributions to $\kappa$
\begin{equation}
\kappa = \kappa_{anapole} + \kappa_{Z^0} + \kappa_{Q_W}
\end{equation}
arise from the nuclear anapole moment, the vector(electron)-axial(nucleus)
tree-level $Z^0$ exchange, and a term generated by the combined
effects of the coherent $Z^0$ and magnetic hyperfine interactions
between the electrons and the nucleus.
In the theory discussions below we will treat the Tl constraint
as one on the principal isotope, $^{205}$Tl (70.5\%).
The other isotope, $^{203}$Tl (29.5\%), differs in structure
only by a pair of neutrons, and thus should have very similar
properties.

The tree-level vector(electron)-axial(nucleus) $Z^0$ interaction
generates a contribution
\begin{equation}
\kappa_{Z^0} = - {g_A \over 2} (1 - 4 \sin^2 \theta_W) 
{< I || \sum^A_{i=1} \sigma(i) \tau_3(i) || I> \over
<I || \hat{I} || I>},
\end{equation}
where $g_A = 1.26$ is the axial-vector coupling and $\sin^2 \theta_W =
0.223$.  To get a rough feel for this contribution, we can evaluate
the nuclear matrix element in the extreme single-particle limit.
$^{133}$Cs would then be described as a unpaired $1g_{7/2}$ proton
outside a closed core, while $^{205}$Tl 
corresponds to an unpaired $3s_{1/2}$ proton.  In this limit
\begin{equation}
\kappa_{Z^0}^{single~particle} = -(-1)^{I-\ell-m_t}
{g_A \over 2 \ell + 1} (1-4 \sin^2 \theta_W)
\end{equation}
where $\ell$ is the single-particle orbital angular momentum
and $m_t$ the $z$-component of isospin (with $m_t$ = 1/2 denoting
a proton).  Thus the single-particle estimates for $^{133}$Cs
and $^{205}$Tl are 0.0151 and -0.136, respectively.
Nuclear models of various types have been employed to try 
to estimate the effects of strong correlations in quenching
the Gamow-Teller matrix element in Eq. (24) from its single-
particle value~\cite{haxton89,hlrm,telitsin,auerbach}.  We will employ
the results of a recent large-basis shell model (SM) study~\cite{hlrm}
here, which yielded
\begin{eqnarray}
\kappa^{SM}_{Z^0}(^{133}\mathrm{Cs}) &=& 0.0140 \nonumber \\
\kappa^{SM}_{Z^0}(^{205}\mathrm{Tl}) &=& -0.127.
\end{eqnarray}
(We will describe these calculations in more detail later.)
In addition, one-loop standard model electroweak radiative corrections,
which we do not include in the numerical results below,
modify the tree-level expression in Eq. (24) somewhat, reducing the
isovector contribution and inducing a small isoscalar
amplitude~\cite{zhu}.

A second contribution to $\kappa$ is generated by the combined
effects of the coherent vector $Z^0$ coupling to the nucleus,
proportional to the weak charge $Q_W$, and the magnetic hyperfine
interaction~\cite{flambaum85}.  From the measured nuclear weak
charge and magnetic moment Bouchiat and Piketty find~\cite{bouchiat91}
\begin{eqnarray}
\kappa_{Q_W}(^{133}\mathrm{Cs}) &=& 0.0078 \nonumber \\
\kappa_{Q_W}(^{205}\mathrm{Tl}) &=& 0.044.
\end{eqnarray}
Thus the experimental values for the anapole contributions to
$\kappa$ are obtained by subtracting the results of Eqs. (26)
and (27) from Eqs. (22), yielding
\begin{eqnarray}
\kappa_{anapole}(^{133}\mathrm{Cs}) &=& 0.090 \pm 0.016 \nonumber \\
\kappa_{anapole}(^{205}\mathrm{Tl}) &=& 0.376 \pm 0.400.
\end{eqnarray}

\subsection{The Hadronic Weak Interaction}

The various mechanisms operating within the nucleus to generate
$\kappa_{anapole}$ arise from the hadronic weak interaction.
As we noted in the introduction, the only practical strategy
for studying the effects of $Z^0$ exchange between quarks is
the investigation of the PNC $NN$ interaction.
As anapole moments are now measureable in high precision atomic
PNC experiments, they now become a part of that strategy.

The low-energy hadronic weak interaction can be described by a
phenomenological current-current Lagrangian~\cite{adelberger}
\begin{equation}
L = {G_F \over \sqrt{2}} (J_W^\dagger J_W + J_Z^\dagger J_Z) + h.c.
\end{equation}
where $J_W$ and $J_Z$ are the charged and neutral weak currents, respectively.
If one considers only the light-quark ($u,d,s$) contributions
to these currents, then $J_W$ has two components
\begin{equation}
J_W = \cos \theta_C J_W^0 + \sin \theta_C J_W^1
\end{equation}
where $\sin \theta_C \sim 0.22$.  The current $J_W^0$ drives the
$u \rightarrow d$ transition and transforms as $\Delta I = 1$,
$\Delta S =0$, while $J_W^1$ drives the $u \rightarrow s$ 
transition and transforms as $\Delta I = 1/2$, $\Delta S = -1$.
(Here $I$ denotes isospin and $S$ strangeness.)  The neutral
current also has two components, $J_Z^0$ and $J_Z^1$, which 
transform as $\Delta I = 0,$ $\Delta S = 0$ and $\Delta I = 1$,
$\Delta S = 0$, respectively.  The $\Delta S =0$ $NN$ 
interaction is then governed by the following piece of Eq. (29)
\begin{eqnarray}
L^{\Delta S =0} &=& {G_F \over \sqrt{2}} [\cos^2 \theta_C (J^0_W)^\dagger J_W^0
+ \sin^2 \theta_C (J_W^1)^\dagger J_W^1 \nonumber \\
&+& (J_Z^0)^\dagger J_Z^0 + (J_Z^1)^\dagger J_Z^1
+ (J_Z^0)^\dagger J_Z^1 + (J_Z^1)^\dagger J_Z^0] + h.c.
\end{eqnarray}

An important aspect of this equation is its isospin content.
The symmetric product of two $J_W^0$ $(\Delta I =1)$ currents
transform as $\Delta I$ = 0 and 2, while the symmetric product of
two $J_W^1$ $(\Delta I = 1/2)$ currents transforms as $\Delta I$ =1.
Thus the $\Delta I = 1$ component of the charged current weak
$NN$ interaction is suppressed by $\tan^2 \theta_C$ compared to
the $\Delta I = 0,2$ components, while the neutral current 
$\Delta I = 1$ contribution is unsuppressed.  It follows that 
the neutral current should dominate this isospin channel.

The main goals of the field have been to isolate this neutral
current and, more generally, to understand the mechanism by which the weak 
force is communicated over the relatively long distances
characterizing low-energy $NN$ scattering or $NN$ interactions
within the nucleus.  While the standard model specifies the
elementary couplings of the weak bosons to quarks, these vertices
are dressed by the strong interaction to form couplings 
between physical particles, such as mesons to nucleons.
We know, from the $\Delta I = 1/2$ rule in strangeness-changing
hadronic weak interactions (the strong enhancement of 
the ratio of $\Delta I = 1/2$ to 3/2 amplitudes), that strong
interaction effects can substantially alter couplings from 
their underlying bare values.

The low-energy $NN$ weak interaction is conventionally described in a one-meson-exchange
model, where one meson-nucleon vertex is weak and the other
strong: this long-distance mechanism dominates at nuclear
densities.  Six weak couplings, $f_\pi$, $h_\rho^0$, $h_\rho^1$,
$h_\rho^2$, $h_\omega^0$, and $h_\omega^1$, characterize the
strengths of the isovector $\pi$, isoscalar/isovector/isotensor $\rho$, and 
isoscalar/isovector $\omega$ weak meson-nucleon couplings~\cite{DDH}.
This model is not as restrictive as it may first appear.
First, CP invariance forbids any coupling between neutral
$J=0$ mesons and on-shell nucleons, eliminating a number of 
candidate interactions~\cite{barton}.  Second, the most general 
low-energy PNC interaction contains only five
independent $S-P$ amplitudes.  From this perspective, the description in 
terms of $\pi$, $\rho$, and $\omega$ exchange can be viewed as an
effective theory, valid at momentum scales much below the inverse range of 
the vector mesons.  At low momentum the detailed short-range behavior
of the $S-P$ potentials is not resolvable: thus one could characterize
the short-range weak interaction by five contact interactions 
corresponding to these independent $S-P$ amplitudes, supplemented
by long-range $\pi$ exchange.  The six meson-nucleon weak
couplings allow one to mimic these six degrees of freedom.

Denoting the isoscalar strong meson-nucleon couplings by
$g_{\pi NN}$, $g_\rho$, and $g_\omega$, the resulting $NN$
PNC potential is~\cite{adelberger}
\begin{eqnarray}
V^{PNC}(\vec{r}) &=& {i g_{\pi NN} f_\pi \over \sqrt{32} M}
[\vec{\tau}(1) \times \vec{\tau}(2)]_3 [\vec{\sigma}(1)+
\vec{\sigma}(2)] \cdot \vec{\mathrm{u}}(\vec{r}) \nonumber \\
&-& {1 \over 2 M} \left[ g_\rho \left( h_\rho^0 \vec{\tau}(1)
\cdot \vec{\tau}(2) + {h_\rho^1 \over 2} [\tau_3(1)+\tau_3(2)] \right. \right. \nonumber \\
 &+& \left. {h_\rho^2 \over 2 \sqrt{6}} [3 \tau_3(1)\tau_3(2) -
\vec{\tau}(1) \cdot \vec{\tau}(2)] \right) \nonumber \\
&\times& \left((1+\mu_{\mathrm{v}}) i [\vec{\sigma}(1) \times \vec{\sigma}(2)]
\cdot \vec{\mathrm{u}}_\rho(\vec{r}) +
[\vec{\sigma}(1)-\vec{\sigma}(2)] \cdot \vec{\mathrm{v}}_\rho(\vec{r}) \right) \nonumber \\
&+& g_\omega \left( h_\omega^0 + {h_\omega^1 \over 2} [\tau_3(1)+
\tau_3(2)] \right) \nonumber \\
&\times& \left( (1+\mu_{\mathrm{s}}) i [\vec{\sigma}(1) \times \vec{\sigma}(2)]
\cdot \vec{\mathrm{u}}_\omega(\vec{r}) + [\vec{\sigma}(1)-
\vec{\sigma}(2)] \cdot \vec{\mathrm{v}}_\omega(\vec{r}) \right) \nonumber \\
&+& \left. {1 \over 2} [\tau_3(1)-\tau_3(2)] [\vec{\sigma}(1)+\vec{\sigma}(2)]
\cdot [g_\omega h_\omega^1 \vec{\mathrm{v}}_\omega(\vec{r}) -
g_\rho h_\rho^1 \vec{\mathrm{v}}_\rho(\vec{r})] \right]
\end{eqnarray}
where $\vec{r} = \vec{r}_1-\vec{r}_2$, $\vec{\mathrm{u}}$ =
$[\vec{p},e^{-m r}/4 \pi r ]$, $\vec{\mathrm{v}}$ =
$\{ \vec{p},e^{-mr}/4 \pi r \}$, and 
$\vec{p}$ = $\vec{p}_1$ - $\vec{p}_2$,
with $\vec{r}_1$ and $\vec{p}_1$ the coordinate and
momentum of nucleon 1.  This expression is usually 
evaluated assuming vector dominance, which fixes the strong
scalar and vector magnetic moments, $\mu_s$ = -0.12 and
$\mu_v$ = 3.70. 

``Best values'' and broad ``reasonable ranges'' for the 
weak meson-nucleon couplings were defined some time ago
by Donoghue, Desplanques, and Holstein~\cite{DDH} (DDH), who
deduced standard-model estimates for these
vertices by using techniques like factorization,
the quark model, and current algebra and sum rule 
methods.   The broad ``reasonable ranges'' reflect
the large degree of uncertainty implicit in such
approximate tools, as well as the potential consequences
of missing physics, such as strange-quark amplitudes~\cite{savage}.
Nevertheless, the DDH results in Table~\ref{table2} have provided
experimentalists with benchmarks for PNC experiments.

\begin{table}
\caption{Weak meson-nucleon coupling ``best values'' and
``reasonable ranges'' (in units of $10^{-6}$) from the standard-model calculations of
Desplanques, Donoghue, and Holstein}
\label{table2}
\begin{center}
\begin{tabular}{ccc}
\hline \hline
Coupling & ``Best Value'' & ``Reasonable Range'' \\
\hline
$f_\pi$ & 0.46 & 0.00$\rightarrow$1.14 \\
$h_\rho^0$ & -1.14 & 1.14$\rightarrow$-3.08 \\
$h_\rho^1$ & -0.019 & -0.038$\rightarrow$0.00 \\
$h_\rho^2$ & -0.95 & -0.76$\rightarrow$-1.10 \\
$h_\omega^0$ & -0.19 & 0.57$\rightarrow$-1.03 \\
$h_\omega^1$ & -0.11 & -0.19$\rightarrow$-0.08 \\
\hline
\end{tabular}
\end{center}
\end{table}

In an ideal world one would determine the low-energy
$NN$ $S-P$ amplitudes, or equivalently the six 
weak meson-nucleon couplings, by a series of $NN$
scattering experiments.  Such experiments 
require measurements of asymmetries $\sim$ $10^{-7}$, which is the
natural scale for the ratio of weak and strong amplitudes,
$4 \pi G_F m_\pi^2 /g_{\pi NN}^2$.  Only a single $NN$
measurement, the longitudinal analyzing power $A_z$
for $\vec{p}+p$, has been successful~\cite{balzer,potter,ramsay}.  (Experiments have been
done at 13.6, 45, and 221 MeV.)  These results have been
supplemented by a number of PNC measurements in nuclear
systems, where accidental degeneracies between pairs of
opposite-parity states
can produce, in some cases, large enhancements in the 
PNC signal.  The experiments include $A_z$ for $\vec{p} + \alpha$
at 46 MeV~\cite{lang}, the circular polarization $P_\gamma$ of the $\gamma$-ray
emitted from the 1081 keV state in $^{18}$F~\cite{bini}, and $A_\gamma$ for
the decay of the 110 keV state in polarized $^{19}$F~\cite{elsener}.
It is widely agreed that these experiments can be interpretted
relatively free of nuclear structure uncertainties: the systems 
are either few-body, where quasi-exact structure calculations 
can be done, or involve special nuclei where the PNC mixing
matrix elements can be calibrated from axial-charge $\beta$ 
decay~\cite{haxton81}.  An anaysis of these results, which have been in hand 
for some time, suggests that the isoscalar PNC $NN$ interaction --
which is dominated by $\rho$ and $\omega$ exchange -- is 
comparable to or slightly stronger than the DDH ``best value,''
while the isovector interaction -- dominated by $\pi$ exchange --
is significantly weaker ($\lsim$ 1/3)~\cite{adelberger}.  As the isovector channel
is expected to be enhanced by neutral currents, there has been 
great interest in confirming this result.  The $^{133}$Cs anapole
result thus provides a possible crosscheck on this tentative 
conclusion that an isospin anomaly, superficially like the
$\Delta I = 1/2$ rule in flavor-changing decays, may exist in
the $\Delta S = 0$ weak $NN$ interaction.

\subsection{Weak Meson-Nucleon Couplings and the Anapole Moment}
  
We denoted in the introduction that an anapole moment -- unlike
the magnetic moment -- of elementary fermions is not a gauge 
invariant quantity.  The very nice discussion of this point 
offered by Musolf~\cite{musolfphd} ends with a very straightforward
explanation: because PNC corrections to the electromagnetic
current couple only to virtual photons, the amplitude for PNC
photon emission is not a physical amplitude and thus need
not be gauge independent.  However the long-distance contributions
to the anapole moment in the nucleus -- the meson cloud contributions
to the nucleon anapole moment and the many-body contributions due
to the PNC $NN$ interaction and associated exchange currents --
are both the dominant contribution to the nuclear anapole
moment and separately gauge invariant~\cite{haxton89}.  These 
contributions, associated with the weak meson-nucleon couplings,
will now be discussed.

To evaluate the nuclear anapole moment in the context of some
model of hadronic PNC one must take the ground-state expectation
value of the operator in Eq. (9).  If one works to first-order
in the weak interaction, two types of terms contribute to the
current matrix element.  First are terms corresponding to the
PNC weak contributions to $J_{1 \lambda}$.  In the
context of the weak meson-nucleon couplings, these include
both one-body terms -- mesonic loop corrections to the ordinary
electromagnetic current involving one weak and one strong
vertex, as illustrated in Fig. 7a -- and exchange currents
(Fig. 7b), where the weak and strong vertices attach to
different nucleons.  We stress that if the anapole operator
of Eq. (9) is employed, the exchange currents are model 
dependent in the sense that all constraints imposed by
current conservation are already explicitly enforced. 
The argument sometimes heard that minimal substitution
in simpler, independent-particle treatments somehow accounts
for the exchange currents is incorrect: once Eq. (9) is 
employed, the surviving exchange currents depend on aspects
of the $NN$ interaction not constrained by current conservation.
  
\begin{figure}[!ht]
\psfig{bbllx=50pt,bblly=0pt,bburx=480pt,bbury=400pt,figure=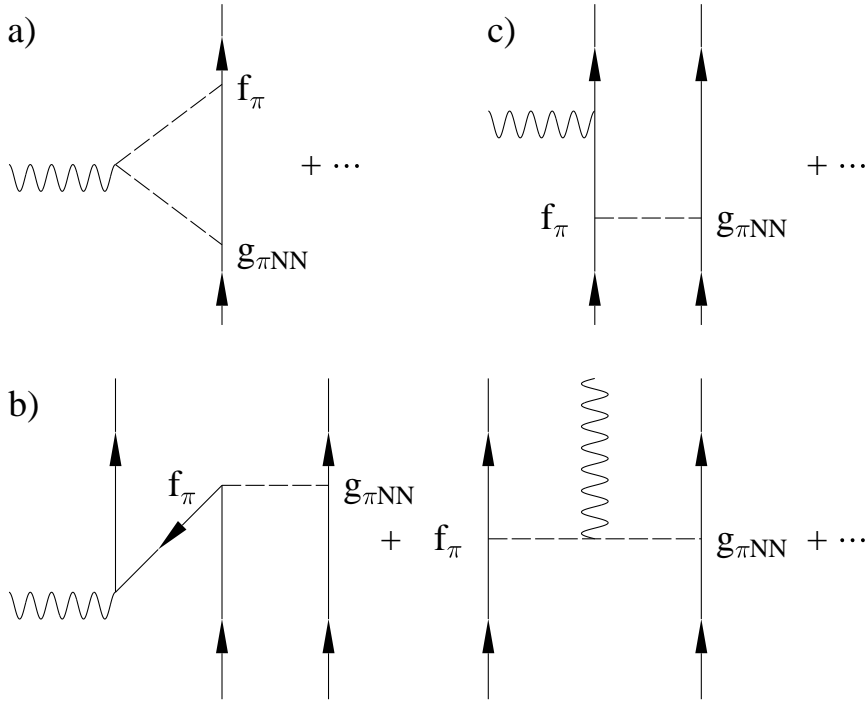,height=10cm}
\caption{The one-body (a), exchange current (b), and nuclear
polarization (c) contributions to the nuclear anapole moment.}
\end{figure}
 
The second class of contributions comes from evaluating 
Eq. (9) for the ordinary, parity-conserving $J_{1 \lambda}^{em}$,
which contributes through PNC admixtures in the nuclear wave
function, as illustrated in Fig. 7c.  The polarization 
term, which depends on the matrix elements of $V^{PNC}$ between the ground
state and a complete set of excited opposite-parity states, can be
enhanced if there exist opposite-parity states very near the
ground state: the admixing is clearly inversely proportional
to the energy difference~\cite{doublet}.  However, in the absence of
such ``accidental'' degeneracies, one
expects the mixing to be dominated by the giant resonances,
the collective states at $\sim 15-20$ MeV in heavy nuclei that account for
most of the $E1$ and other first-forbidden nuclear response.

We discuss each of these contributions below, depending rather
heavily on the recent work of Ref.~\cite{hlrm}.  In Ref.~\cite{hlrm}
an effort was made to model the various contributions to the 
anapole moment using the formalism that has become standard in
other studies of hadronic PNC: the DDH weak-meson couplings,
the two-body $V^{PNC}$ based on those couplings, large-basis
SM nuclear wave functions, and a standard correlation
function to modified the short-distance behavior of the 
SM two-nucleon density.  Thus this treatment will
allow us, in Section 5, to compare anapole, $NN$, and nuclear
constraints on hadronic PNC on an equal footing.
  
\noindent
{\it a) Nucleon anapole moment.}  The
one-body PNC electromagnetic current derived from meson loop
corrections (e.g., Fig. 7a) yields in the nonelativistic limit
\begin{equation}
\hat{A}_{1 \lambda}^{nucleonic} = \sum_{i=1}^A \left[a_s(0) + a_v(0) \tau_3(i)\right] \sigma_{1 \lambda}(i).
\end{equation}
That is, this term is the sum of the anapole moments of the
individual nucleons.  As 
the contribution of spin-paired core nucleons cancel, one expects
this term to depend on the unpaired valence nucleon in the odd-$A$
nuclei of interest.  Thus the one-body contribution should be roughly
independent of $A$ though -- like the closely related example 
of magnetic moments -- its value will depend on the 
shells occupied by the valence nucleon.  Some time ago the
pion contribution to $a_s(0)$ and
$a_v(0)$ was evaluated~\cite{haxton89}, yielding results
proportional to $e f_\pi g_{\pi NN}$.  The isoscalar coupling $a_s(0)$
proved to be about four times larger than $a_v(0)$.  Recent extensions of this work  
have included the full set of one-loop contributions involving
the DDH vector meson PNC couplings, using the framework of heavy baryon
chiral perturbation theory and retaining contributions through $O(1/\Lambda_\chi^2)$,
where $\Lambda_\chi = 4 \pi F_\pi \sim 1$ GeV is the scale of chiral symmetry breaking \cite{zhu}.
The contributions due to $f_\pi$ are consistent with the earlier
work -- the new $a_s(0)$ is about 1.3 times larger, while $a_v(0)$
is zero in this order -- so that the dominance of $a_s(0)$ in
Ref.~\cite{haxton89} was is explained as a consequence of
heavy baryon chiral perturbation theory.
The addition in Ref.~\cite{zhu} of the heavy meson contributions
greatly enhances $a_v(0)$.  An evaluation with DDH best value couplings
yields $a_v(0) \sim 7a_s(0)$.    
Thus the inclusion of heavy meson contributions substantially enhances
the one-body anapole terms and alters the isospin character,
generating opposite signs for the proton and neutron anapole moments.

This then determines $\hat{A}_{1 \lambda}^{one body}$.  The SM
studies of Ref.~\cite{hlrm}, which we will describe in
more detail below, give matrix elements
$\langle I || \sum_{i=1}^A \sigma(i) || I \rangle$ = -2.37 and 2.53 and 
$\langle I || \sum_{i=1}^A \sigma(i) \tau_3(i) || I \rangle$ = -2.30 and 2.28 
for $^{133}$Cs and $^{205}$Tl, respectively.
This leads to the nucleonic anapole contributions shown in
Table~\ref{table3}.
  
\begin{table}
\caption{Shell model estimates of the anapole
matrix element $\langle I || A_1 || I \rangle/e$,
expressed as cofficients times the indicated weak couplings}
\label{table3}
\begin{center}
\begin{tabular}{ccrrrrrr}
\hline \hline
Nucleus&Source&$f_\pi$&$h_\rho^0$&$h_\rho^1$&$h_\rho^2$&$h_\omega^0$&$h_\omega^1$ \\
\hline
$^{133}$Cs&nucleonic&0.59&0.87&0.90&0.36&0.28&0.29 \\
&ex. cur.&8.58&0.02&0.11&0.06&-0.57&-0.57 \\
&polariz.&51.57&-16.67&-4.88&-0.06&-9.79&-4.59 \\
&total&60.74&-15.78&-3.87&0.36&-10.09&-4.87 \\
$^{205}$Tl&nucleonic&-0.63&-0.86&-0.96&-0.35&-0.29&-0.29 \\
&ex. cur.&-3.54&-0.01&-0.06&-0.03&0.28&0.28 \\
&polariz.&-13.86&4.63&1.34&0.08&2.77&1.27 \\
&total&-18.03&3.76&0.33&-0.30&2.76&1.26 \\
\hline
\end{tabular}
\end{center}
\end{table}

\noindent
{\it b) Exchange currents.}  Two-body PNC currents arise from
meson exchange diagrams where the photon couples to the meson
``in flight'' (transition currents), or where the photon and
either the PC or PNC meson-nucleon vertex creates and annihilates
an $N \bar{N}$ pair (pair currents), as illustrated
in Fig. 7b.  The pionic exchange currents, which have the longest
range, were evaluated in Ref.~\cite{haxton89}.  The explicit form of
the pion contribution to $A_{1 \lambda}^{ex~cur}$ is given 
there.  Rather than quote the result, which is a bit complicated,
we instead give the one-body reduction of the pair current contribution to that operator, which
illustrates the underlying physics much more clearly.  
  
If one views a nucleus as a single-particle outside of a closed
core, a two-body operator can be replaced by an 
equivalent one-body effective operator
\begin{equation}
<\alpha | A^{eff}_{1 \lambda} | \beta > =
\sum_{\delta < F} < \alpha \delta | A_{1 \lambda}^{ex~cur} | \beta \delta - \delta \beta>,
\end{equation}
where the sum is taken over the nuclear core.  Completing this
sum in Fermi gas model, assuming a spin-symmetric but
isospin-asymmetric core, yields the pionic pair current effective
operator
\begin{eqnarray}
A^{eff, N \bar{N}}_{1 \lambda} &=& 2.74 a_s^\pi(0) {M \over m_\pi^2}
\sum_{i=1}^A \rho(r_i) r_i^2 (\sigma_{1 \lambda}(i) + \sqrt{2 \pi}
[Y_2(\hat{r}_i) \otimes \sigma(i)]_{1 \lambda}) \nonumber \\
&\times& \left( {Z \over A} \omega_Z^\pi (1 - {2 \over 3} \tau_3(i))
+ {N \over A} \omega_N^\pi (1 + {2 \over 3} \tau_3(i)) \right),
\end{eqnarray}
where $\rho(r_i)$ is the nuclear density operator and
$\omega_Z^\pi$ ($\omega_N^\pi$) a proton (neutron) Fermi-gas
response function that depends on $k_i/k_F$, the nucleon 
momentum as a fraction of the Fermi momentum.  The $\omega$s
vary only gently, ranging from 0.33 to 0.19 as $k_i/k_F$
increases from 0 to 1.  Thus a suitable average value is
$\sim$ 0.25.
  
The overall strength of $A^{eff, N \bar{N}}_{1 \lambda}$ is given in terms of the pionic contribution
to the single-particle anapole moment
\begin{equation}
a_s^\pi(0) \sim -1.6 {e f_\pi g_{\pi NN} \over 8 \sqrt{2} \pi^2}
\end{equation}
to allow an easy comparison with the one-body current.  Using a
nuclear density of 0.195/fm$^3$, one finds that the pair
isoscalar $\pi$ exchange current then scales as $\sim 0.9 A^{2/3} a_s^\pi(0)$.
Unlike the nucleonic contribution, the exchange current 
contribution grows as $A^{2/3}$.  Clearly it will increasingly
dominate over the nucleonic contribution as $A$ increases.
The isovector effective operator is smaller by a factor of
$2(Z-N)/3A$, reflecting a cancellation between contributions
from core protons and neutrons.

The calculations of Ref.~\cite{haxton89} employed the full two-body
form for the pionic currents, evaluating these from the SM
two-body density matrices.  A short-range correlation function
was introduced to mimic the effects of hard-core correlations 
on the density matrix.  The extension
to include the $\rho$ and $\omega$ PNC couplings is a formidable
task requiring evaluation of the $\rho$ and $\omega$ pair currents
and the $\rho \rho \gamma$ and $\rho \pi \gamma$ currents.
This calculation was only recently completed~\cite{hlrm}.
The $\rho \rho \gamma$ and $\rho \pi \gamma$ transition currents
and the component of the $\omega$ pair current
where the photon and PNC $\omega$ couplings are on different 
nucleon legs were found to be negligible, well below 1\% of the dominant
$\pi$ currents; the remaining heavy-meson currents are more
important but, as can be seen in Table~\ref{table3}, still are
suppressed relative to the pionic currents.  The tabulated 
results were obtained from the SM calculations described below.
  
\noindent
{\it c) Nuclear polarization contribution.}  The nuclear polarization
contribution (Fig. 7c) to the anapole moment is given by 
\begin{equation}
\sum_n {\langle I || \hat{A}^{em}_1 || n \rangle \langle n | H^{PNC} | I \rangle
\over E_{gs} - E_n } + h.c. 
\end{equation}
where $\hat{A}^{em}_1$ is generated from inserting the ordinary electromagnetic current
operator into Eq. (9), $| I \rangle$ is a ground state of good parity, $H^{PNC}$
is the PNC NN interaction, and the sum extends over a complete
set of nuclear states $n$ of angular momentum $I$ and opposite parity.
The extended Siegert's theorem again determines the form of
$\hat{A}_1^{em}$~\cite{haxton89}.
  
Dmitriev, Khriplovich, and Telitsin~\cite{dkt} have provided a series of
estimates of the polarization contribution in the context 
of single particle models, using a one-body effective $V^{PNC}$.
The models include one of uniform density and ones based
on the harmonic oscillator and Woods-Saxon potentials.
The former yields an anapole operator that is explicitly proportional
to the nuclear density -- a quantity approximately constant
in heavy nuclei because of nuclear saturation -- and grows
as $A^{2/3}$, like the exchange current contribution.  
Numerically, however, the polarization contribution is about
a factor of five larger than the exchange currents: it is the dominant contribution 
to axial PNC in heavy nuclei.

For the present analysis we need a treatment of polarization
contribution that begins with the DDH form of $V^{PNC}$.
This presents two nuclear structure challenges.
The first is the construction of a reasonable model for the
ground state.  The SM calculations of Refs.~\cite{haxton89,hlrm}
for $^{133}$Cs were done in the canonical space between magic shells 50 and 82,
$1g_{7/2}-2d_{5/2}-1h_{11/2}-3s_{1/2}-2d_{3/2}$.   
The protons were restricted to the first two of these
shells and neutron holes to the last three, producing an m-scheme
basis of about 200,000.  
$^{205}$Tl is described as a proton hole in the   
orbits immediate below the Z=82 closed shell ($3s_{1/2}-2d_{3/2}-2d_{5/2}$)
coupled to two neutron holes in valence neutron space between
magic numbers 126 and 82 ($3p_{1/2}-2f_{5/2}-3p_{3/2}-1i_{13/2}-2f_{7/2}-1h_{9/2}$).

The second challenge is the completion of the intermediate state sum.
Direct or moments methods~\cite{haxton89} are impractical because of the dimensions of the spaces.
However, because no nonzero $E1$ transitions exist among the 
valence orbitals, an alternative of completing the sum by closure,
after replacing $1/E_n$ by an average value $\langle 1/E \rangle$,
is quite attractive: the resulting product of the one-body operator
$A^{em}_1$ and two-body $V^{PNC}$ contracts to a two-body operator, so that
only the two-body ground state density matrix is needed.
The closure aproximation can be considered an identity if one
knows the correct $\langle 1/E \rangle$: in practical terms,
this means demonstrating that it can be related reliably to 
some known quantity, with the position of the $E1$ giant
resonance being the obvious candidate.  A series of SM
calculations has been completed for a series of light nuclei, where both
the $1/E-$ and non-energy-weighted sums could be done~\cite{hlrm}.
It was found that there was a consistent relationship between the
closure energy and the giant dipole position, though the three
isospin contributions to $V^{PNC}$ must be treated separately.
The average excitation energies found (as fractions of the dipole energy) are
$0.604 \pm 0.056~(h_\rho^0, h_\omega^0)$, 
$0.899 \pm 0.090~(f_\pi)$, and $1.28 \pm 0.14~(h_\rho^2)$.
The larger $\langle 1/E \rangle$ for $h_\rho^0$ and $h_\omega^0$ enhances these
contributions to the anapole polarizability.  
The assumption that these relations also hold~\cite{doublet}
in the heavier nuclei $^{133}$Cs and $^{205}$Tl
then fixes the appropriate average excitation energies for
these nuclei.  From the SM two body density matrix, 
the contracted two-body effective operator that results
from the closure approximation (this includes the effects
of a short-range correlation function on the two-nucleon
matrix elements of $V^{PNC}$), and these excitation
energies, one obtains the SM polarization results in Table~\ref{table3}.

Note that there are situations -- chance ground-state parity 
doublets or nuclear octupole deformation -- that can lead to
enhanced polarizabilities.  While such enhancement is unexpected
in $^{133}$Cs as the nucleus is approximately spherical, 
enhancements have been discussed in connection with other nuclei
of interest because of anapole or electric dipole moments~\cite{doublet}.

\section{WEAK COUPLING CONSTRAINTS}

From Table~\ref{table3} one has the matrix element of $A_{1 \lambda}$
as an expansion in terms of the DDH weak couplings.  By Eq. (12),
one has a corresponding expression for $\kappa_{anapole}$,
which can be compared to the experimental constraints 
(Eq. (28)).

The constraints from $^{133}$Cs and Tl supplement those
available from experiments in $\vec{p}+p$ scattering and 
in nuclear systems.  While a larger set of results exists,
only some of these are generally regarded as being
reliably interpretable~\cite{adelberger}.  Rather precise
measurements of the longitudinal analyzing power $A_z$
for $\vec{p}+p$ have been made at 13.6 and 45 MeV, and a
preliminary result at 221 MeV (where only $\rho$ exchange
contributes) is now available.  $A_z$ has also been measured
for $\vec{p}+\alpha$ at 46 MeV.  There are also two important
constraints from nuclei in which observables associated with
nearly degenerate parity doublets have been measured.
In each case the nuclear matrix element involved in the
mixing has been determined from axial-charge $\beta$ 
decay~\cite{adelberger,haxton81}, so that little nuclear
structure uncertainty remains.  The observables are the
circular polarization $P_\gamma$ of the $\gamma$-ray emitted
in the decay of the 1081 keV state in $^{18}$F and 
angular asymmetry $A_\gamma$ for the decay of the 110 keV
state in polarized $^{19}$F.
  
\begin{figure}[!ht]
\psfig{bbllx=0pt,bblly=90pt,bburx=540pt,bbury=690pt,figure=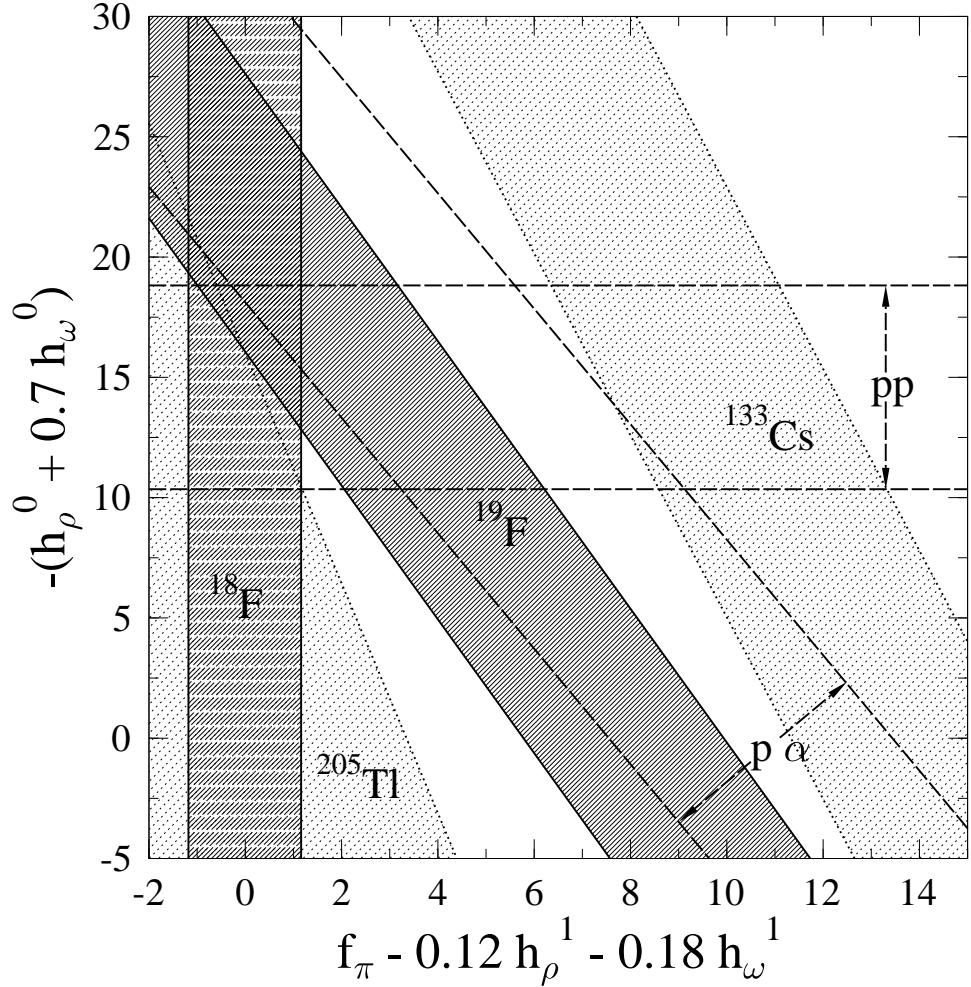,height=13.5cm}
\caption{Constraints on the PNC meson couplings ($\times 10^7$)
that follow from the results in Table~\ref{table4}.  The error bands are
one standard deviation.  The illustrated region contains all
of the DDH ``reasonable ranges'' for the indicated parameters.}
\end{figure}
  
A summary of PNC constraints is presented in Table~\ref{table4} and Fig. 8.
Although the PNC parameter space is six-dimensional, two coupling
constant combinations, $f_\pi-0.12h_\rho^1-0.18h_\omega^1$ and
$h_\rho^0+0.7h_\omega^0$, dominate the observables, as Table~\ref{table4}
illustrates.  The $1\sigma$ error bands of Fig. 8 are generated
from the experimental uncertainties, broadened somewhat by allowing
uncorrelated variations in the parameters in the last four columns
of Table~\ref{table4} over the DDH broad ``reasonable ranges.''  
Note that only a fraction of the region allowed by the Seattle Tl
constraint is shown: the total width of the Tl band is an order of
magnitude broader than the width of the Cs allowed band, with most
of the Tl allowed region lying outside the DDH ``reasonable ranges''
(i.e., in the region of negative $f_\pi$ and positive $h_\rho^0+0.7h_\omega^0$).
The corresponding Oxford Tl band, which is not illustrated, includes
almost all of the parameter space in Fig. 8, as well as a substantial 
region (to the lower left of Fig. 8) outside the bounds of the figure.

The weak coupling ranges covered by Fig. 8 correspond roughly to
the DDH broad ``reasonable ranges.''  Thus the anapole constraints
are not inconsistent with the most general theory constraints.
However in detail, the
pattern is disconcerting.  Before the anapole results are included,
the indicated solution is a small $f_\pi$ and an isoscalar coupling 
somewhat larger, but consistent with, the DDH best value, 
$-(h_\rho^0 + 0.7h_\omega^0)^{DDH}_{b.v.} \sim 12.7$.  The
anapole results agree poorly with the indicated solution,
as well as with each other.  Although the Seattle Tl measurement is consistent
with zero, it favors a positive anapole moment, while the theory
prediction is decidedly negative, given existing PNC constraints.
The Cs result tests a combination of PNC couplings quite similar
to those measured in $A_\gamma(^{19}$F) and in $A_z^{p \alpha}$,
but favors larger values.

\begin{table}
\caption{PNC observables and corresponding theoretical predictions,
decomposed into the designated weak-coupling combinations,
with $\tilde{f}_\pi = f_\pi-0.12h_\rho^1-0.18h_\omega^1$ and
$\tilde{h}^0 = h_\rho^0+0.7h_\omega^0$}
\label{table4}
\begin{center}
\begin{tabular}{cccccccc}
\hline \hline
Observable&Exp.($\times 10^{7}$)&$\tilde{f}_\pi$&$\tilde{h}^0$&$h_\rho^1$&$h_\rho^2$&$h_\omega^0$&$h_\omega^1$ \\
\hline
$A_z^{pp}(13.6)$&-0.93 $\pm$ 0.21&&0.043&0.043&0.017&0.009&0.039 \\
$A_z^{pp}(45)$&-1.57 $\pm$ 0.23&&0.079&0.079&0.032&0.018&0.073 \\
$A_z^{pp}(221)$&prelim.&&-0.030&-0.030&-0.012&0.021& \\
$A_z^{p \alpha}(46)$&-3.34 $\pm$ 0.93&-0.340&0.140&0.006&&-0.039&-0.002 \\
$P_\gamma(^{18}$F)&1200 $\pm$ 3860&4385&&34&&&-44 \\
$A_\gamma(^{19}$F)&-740 $\pm$ 190&-94.2&34.1&-1.1&&-4.5&-0.1 \\
$\langle || A_1 || \rangle/e,$ Cs&800 $\pm$ 140&60.7&-15.8&3.4&0.4&1.0&6.1 \\
$\langle || A_1 || \rangle/e,$ Tl&370 $\pm$ 390&-18.0&3.8&-1.8&-0.3&0.1&-2.0 \\
\hline
\end{tabular}
\end{center}
\end{table}
  
The results plotted in Fig. 8 for $^{133}$Cs are consistent with
those of Flambaum and Murray \cite{flambaum97}, who extract from the anapole moment
an $f_\pi$ about twice the DDH best value, $f_{\pi~b.v.}^{DDH} \sim 4.6$,
and point out that theory can accommodate this.  (The DDH
reasonable range is 0-11.4, in units of $10^{-7}$.)  However, this ignores $P_\gamma(^{18}$F),
a measurement that has been performed by five groups.  The resulting
constraint is almost devoid of theoretical uncertainty
\begin{equation}
-0.6 \lsim f_\pi - 0.11 h_\rho^1 - 0.19 h_\omega^1 \lsim 1.2.
\end{equation}
Allowing $h_\rho^1$ and $h_\omega^1$ to vary throughout their DDH 
reasonable ranges, one finds $-1.0 \lsim f_\pi \lsim 1.1$, clearly
ruling out $f_\pi \sim$ 9.  Fig. 8 illustrates this, as well as the
additional tension between Cs, $p+\alpha$, and $A_\gamma(^{19}$F).

\section{OUTLOOK}

The first conclusion one would draw from Fig. 8 is that 
additional experimental constraints would be helpful.
In the case of nuclear experiments, the situation has been
essentially static for the past 15 years, apart from the 
$\vec{p}+p$ $A_z$ measurement at 221 MeV.  
However, in the next few years results are expected from experiments on the 
PNC spin rotation of polarized slow neutrons in liquid
helium~\cite{heckel} and on $A_\gamma$ in $n + p \rightarrow
d + \gamma$~\cite{snow}.  Thus we could soon have the
information to test whether the small-$f_\pi$ solution
favored by the $\vec{p}+p$ and nuclear experiments is
correct: this is the one consistent solution satisfying the
nuclear constraints.

The anapole situation is less satisfactory: even apart
from the question of consistency with the nuclear constraints, there is
some tension between the $^{133}$Cs and Seattle Tl allowed regions,
though Fig. 8 (which omits the bulk of the Tl allowed region) tends
to exagerate this disagreement: if the Tl band were enlarged to 
$2 \sigma$, it would encompass all of the illustrated region,
including the Cs band.  While the Seattle Tl result favors a sign
opposite theory, its broad error bar allows either sign.  Taken 
together, the Seattle and Oxford results are not incompatible with
any choice of coupling constants with the ``reasonable ranges.''
An improved measurement in
Tl would clearly be helpful, as would any new
anapole measurement involving an odd-neutron nucleus,
which would produce a band in Fig. 8 roughly perpendicular
to that for Cs.  There is an atomic PNC effort underway
using Dy, which has two abundant odd-neutron isotopes,
$^{161}$Dy and $^{163}$Dy~\cite{budker}.  
Another possibility may come from a new atomic PNC technique
using a single trapped Ba$^+$ ion: the much larger 
coherence times and field intensities possible with this
method compensates for the sensitivity loss stemming from
the use of a single atom.  (The statistical accuracy of
traditional methods goes as $\sqrt{N}$, where $N$ is
the number of atoms.)  Ba has two stable odd-neutron
isotopes, $^{135}$Ba and $^{137}$Ba.
As with Cs these attempts will have to reach the sub-1\%
level of precision in order to permit an extraction of the
differential effects due to nuclear spin dependence.  Alternatively
other methods, such as proposals to measure the anapole moment
directly through $E1/M1$ interference in hyperfine
transitions~\cite{gorsh}, could be developed.

The underlying issues are not limited to atomic PNC 
measurements: our understanding of V($e$)-A($N$) interactions affects
the interpretation of electron-nucleus PNC scattering experiments like SAMPLE \cite{beise}, where a
similar discrepancy between theory and experiment exists,
one that could be associated with an incomplete treatment
of PNC effects in the nuclear target.

Figure 8 shows that the jump in precision achieved in the $^{133}$Cs
experiment has now provided a constraint on hadronic PNC that is 
comparable in accuracy to the best of the nuclear constraints.
This is the reason the lack of overlap in the resulting bands is a concern.
The nuclear physics required to analyze the
$^{133}$Cs result is nontrivial~\cite{savagea}, and could
yet prove to be the source of the discrepancies apparent 
in Fig. 8.  The SM calculations described 
in Section 4 depend on a relation between the closure energy
and the giant dipole energy that has been tested only in light nuclei.
It is possible that the systematics for neutron-rich
nuclei might be different.  It is also generally appreciated
that spin-dependent operators tend to be quenched as model
spaces are enlarged to encompass more realistic correlations.
Indeed, phenomenological single-particle treatments of
anapole moments have previously envoked phenomenological
quenching factors~\cite{bouchiat91}.  Our SM results are 
quenched relative to single-particle estimates, and would
likely be further weakened if those calculations could be
enlarged.  Likewise core polarization results from the
RPA study of Dmitriev and Telitsin find substantial 
quenching, relative to single-particle estimates~\cite{telitsin}.
All of these results argue that more realistic treatments
of correlations will tend to reduce matrix element values,
thus requiring still larger values of the weak coupling
constants to fit the experimental results.  Despite this
possibility, it is clear that the degree of effort required
to complete the $^{133}$Cs experiment obligates theorists to
invest similar effort in improving their calculations.
It is an unfortunate situation that the 
results in Table~\ref{table3} come from model calculations:
it is thus difficult to come up with any strategy for quantifying
possible errors other than the comparisons that can be made
when still more ambitious calculations are eventually performed.

This work was support in part by the US Department of 
Energy and by the National Science Foundation.
WH thanks the Miller and Guggenheim Foundations
for support they provided.

\end{document}